\documentclass[english,prl,aps,floatfix,twocolumn,longbibliography,superscriptaddress]{revtex4-1}
\usepackage[T1]{fontenc}
\usepackage[latin9]{inputenc}
\usepackage{color}
\usepackage{rotating}
\usepackage{bm}
\usepackage{amsmath}
\usepackage{amssymb}
\usepackage{graphicx}

\makeatletter

\providecommand{\tabularnewline}{\\}

\usepackage[colorlinks=true,linkcolor=blue,anchorcolor=red,citecolor=blue,urlcolor=blue]{hyperref}
\usepackage{lipsum}

\makeatother

\usepackage{babel}
\begin{document}
\renewcommand{\figurename}{Fig.}
\title{All-electrically tunable networks of Majorana bound states}
\author{Song-Bo Zhang}

\affiliation{Institute for Theoretical Physics and Astrophysics$\text{,}$  University of W\"urzburg, D-97074 W\"urzburg, Germany}
\author{Alessio Calzona}

\affiliation{Institute for Theoretical Physics and Astrophysics$\text{,}$  University of W\"urzburg, D-97074 W\"urzburg, Germany}
\author{Björn Trauzettel}
\affiliation{Institute for Theoretical Physics and Astrophysics$\text{,}$  University of W\"urzburg, D-97074 W\"urzburg, Germany}
\affiliation{W\"urzburg-Dresden Cluster of Excellence ct.qmat, Germany}
\date{\today}

\begin{abstract}
Second-order topological superconductors (SOTSs) host localized Majorana fermions and provide a new platform for topological quantum computation. We propose a remarkable and feasible way to realize networks based on SOTSs which allow to nucleate and braid Majorana bound states (MBSs) in an all-electrical manner without fine-tuning. The proposed setups are scalable in a straightforward way and can accommodate any even number of MBSs. Moreover, the MBSs in the networks allow defining qubits whose states can be initialized and read out by measuring Josephson currents flowing between SOTS islands. Our proposal can be implemented in monolayers of $\text{FeTe}{}_{1-x}\text{Se}_{x}$, monolayers of 1T'-WTe$_2$, and inverted Hg(Cd)Te quantum wells in proximity to conventional superconductors.
\end{abstract}

\maketitle
\textit{Introduction.\textemdash }Second-order topological superconductors
(SOTSs) are characterized by topologically protected midgap bound
states with zero excitation energy and codimension two \citep{Langbehn17PRL,Benalcazar17science,ZDSong17PRL,Benalcazar17PRB,Schindler18SciAdv,QYWang18PRL,ZBYan18PRL,TLiu18PRB,Geier18PRB,Shapourian18PRB,Skurativska20PRR,Tiwari20PRL}.
These midgap states behave like Majorana fermions which constitute
their own anti-particles \citep{Majorana32NC}. They obey non-Abelian
exchange statistics and could find promising applications in topological
quantum computation \citep{Ivanov01PRL,Kitaev03AnPhy,Nayak08RMP,Alicea12PPP,Beenakker13ARCMP,sarma15npj,Elliott15RMP}.
Recently, SOTSs have been predicted in certain candidate systems \citep{QYWang18PRL,ZBYan18PRL,TLiu18PRB,Geier18PRB,CHHsu18PRL, Volpez19PRL,Shapourian18PRB,ZhangSB20PRR,Ghorashi19PRB, RXZhang19PRLb,Plekhanov19PRR,Ahn2019arXiv, Franca19PRB,RXZhang19PRL,Bultinck19PRB,HSu19arXiv,ZBYan19PRL, PanXH19PRL,Laubscher20PRR,Yang20PRR,WYJ19arxiv,XXWu19arXiv190510648W}.
Hence, they provide a feasible platform for implementing topological
quantum gates \cite{YZYou19PRB,Bomantara20PRB,
SBZhang20arXivSOTS}. A few theoretical proposals have been made to explore the exchange of Majorana bound states (MBSs) in SOTSs \citep{XYZhu18PRB,Ezawa19PRB,Pahomi19arXiv,SBZhang20arXivSOTS}.
However, they are restricted to only a single pair of MBSs or require
to locally tune magnetic fields. To define a multidimensional computational
ground-state manifold suitable for implementing non-Abelian quantum
gates, four or more MBSs are required \citep{Bravyi06PRA,Nayak08RMP}.
Moreover, simpler manipulation schemes based on electrical controls
are advantageous in experimental implementation and runtime for quantum
gates.

In this Letter, we propose a novel way to realize electrically tunable
networks of MBSs based on SOTSs. We take full advantage
of the special role played by the sample geometry in SOTSs and conceive
setups whose building blocks consist of isosceles right triangle islands
(IRTIs) of SOTSs. By modulating local gate voltages on the islands,
it is possible to nucleate an arbitrary even number of MBSs and control
their positions on the networks, allowing for non-Abelian braiding.
The magnetic order in our proposal can be uniform. It can, for instance, be realized by in-plane ferromagnetism (FM), antiferromagnetism (AFM), Zeeman
fields, or a mixture of them. Moreover, the qubit states defined
by the MBSs in the network can be initialized and readout, for instance,
by measuring Josephson currents flowing between the SOTS islands.
Importantly, our proposal can be implemented in a variety of candidate
systems, including 1T'-WTe$_2$ monolayers, inverted Hg(Cd)Te quantum wells with proximity-induced
superconductivity and $\text{FeTe}{}_{1-x}\text{Se}_{x}$ monolayers
with intrinsic superconductivity.

\textit{MBSs on open boundaries of SOTSs.}\textemdash We consider two-dimensional SOTSs which are realized by introducing $s$-wave pairing potential in combination with in-plane FM or AFM to quantum spin Hall insulators. The SOTSs can be described by
\begin{eqnarray}
\mathcal{H}({\bf k}) & = & m({\bf k})\tau_{z}\sigma_{z}+A\sin k_{x}s_{z}\sigma_{x}+A\sin k_{y}\tau_{z}\sigma_{y}\nonumber \\
 &  & -\mu\tau_{z}+\Delta_{0}\tau_{y}s_{y}+H_{M}\label{eq:minimal-model}
\end{eqnarray}
in the basis $(c_{a\uparrow},c_{b\uparrow},c_{a\downarrow},c_{b\downarrow},c_{a\uparrow}^{\dagger},c_{b\uparrow}^{\dagger},c_{a\downarrow}^{\dagger},c_{b\downarrow}^{\dagger})$,
where $c_{\sigma s}$ is the fermion operator with orbital (or sublattice)
index $\sigma$$\in$$\{a,b\}$ and spin index $s\in\{\uparrow,\downarrow\}$;
$m({\bf k})=2m\cos k_{x}+2m\cos k_{y}+m_{0}-4m$ with $m_{0}m>0$;
$\mu$ is the chemical potential controllable by external
gates. The Pauli matrices ${\bf s}$, $\bm{{\bf \sigma}}$ and $\bm{\tau}$
act on spin, orbital and Nambu spaces, respectively. $H_{M}$
describes the magnetic order. It can be induced by close proximity
to ferromagnets or antiferromagnets or by applying in-plane magnetic
fields. For concreteness, we focus on the case of FM with strength
$M_{0}$ in  $x$ direction, $H_{M}=M_{0}\tau_{z}s_{x}$ \cite{NoteM}.

The SOTSs feature zero-energy MBSs when open boundary conditions are
enforced. To better understand this, it is instructive to derive a
low-energy effective Hamiltonian on boundaries. We start with
the low-energy limit of $\mathcal{H}({\bf k})$ and consider the SOTSs
in a disk geometry of radius $R$. In the absence of $M_{0}$ and
$\Delta_{0}$, we can find helical states $(\Psi_{e,\uparrow},\Psi_{e,\downarrow}$,
$\Psi_{h,\uparrow},\Psi_{h,\downarrow})$ on the disk boundary.
Using these helical states as a basis and projecting the full Hamiltonian
$\mathcal{H}({\bf k})$ on these states, the boundary Hamiltonian
is constructed as
\begin{equation}
\mathcal{\widetilde{H}}(\varphi) = -Ap_{\varphi}s_{z}+\Delta_{0}\tau_{y}s_{y}- \widetilde{M} e^{-i\tau_zs_z\varphi}s_y - \mu\tau_{z} ,\label{eq:boundary-Hamiltonian}
\end{equation}
where $\varphi$ is the azimuthal coordinate and $p_{\varphi}\equiv-i\partial_{\varphi}/R$
the corresponding momentum defined along the boundary. The boundary states possess effective pairing potential $\Delta_{0}$ and
magnetization $\widetilde{M}=M_{0}\sin\varphi$, as induced from the bulk. When $M_{0}>\bar{\Delta}\equiv(\Delta_{0}^{2}+\mu^{2})^{1/2}$,
we find that the energy bands of Eq.\ (\ref{eq:boundary-Hamiltonian})
change their order at the angles
\begin{equation}
\varphi_{1/4}=\text{\ensuremath{\pm}\ensuremath{\ensuremath{\arcsin}}}(\bar{\Delta}/M_{0}),\ \text{and }\varphi_{2/3}=\varphi_{4/1}+\pi\label{eq:positions}
\end{equation}
along the boundary.
The changes of band order indicate the appearance of four MBSs $\gamma_{i}$
with $i\in\{1,2,3,4\}$, exponentially localized at $\varphi_{i}$.
When they are well separated from each other, the four MBSs are at
zero energy and it is possible to analytically derive their wavefunctions
$\Psi_{i}$\ \citep{SuppInf}. Importantly,
the chemical potential $\mu$ controls the angles $\varphi_{i}$,
according to Eq.\ (\ref{eq:positions}). This enables us to manipulate
the positions of the MBSs, and eventually their fusion and braiding
in an all-electrical manner, as discussed below.

\textit{Fusion properties of MBSs.\textemdash }When two MBSs are brought close together, their wavefunctions start to overlap and their energies
become finite. This process, known as fusion, is mediated by the electron hopping in the SOTSs. According to Eq.\ (\ref{eq:minimal-model}),
the hopping corresponds to the operator $\hat{T}=iA(s_{z}\sigma_{x}+\tau_{z}\sigma_{y})/2+2m\tau_{z}\sigma_{z}$. Thus, the fusion strength between two MBSs, say $\gamma_{i}$ and $\gamma_{j}$, can be estimated as $F_{\gamma_{i}:\gamma_{j}}=|\langle\Psi_{i}|\hat{T}|\Psi_{j}\rangle|$.
On a single island, we find that the fusion strengths $F_{\gamma_{1}:\gamma_{2}}$ and $F_{\gamma_{3}:\gamma_{4}}$ are proportional to $\cos\vartheta$, while $F_{\gamma_{1}:\gamma_{4}}$ and $F_{\gamma_{2}:\gamma_{3}}$ to $\text{sin}\vartheta$, where $\vartheta=\text{arctan}(\mu/\Delta_{0})$. By contrast, the fusion between $\gamma_{1}$ and $\gamma_{3}$ (or $\gamma_{2}$ and $\gamma_{4}$) is strictly forbidden, due to inversion symmetry of the SOTSs \citep{SuppInf}.

The fusion properties become richer when we consider two sets of MBSs
$\{\gamma_{i}\}$ and $\{\gamma_{i}'\}$ (with $i\in\{1,2,3,4\}$)
belonging to two different islands, featuring a finite pairing phase
difference. In this case, when two MBSs from different
islands are brought close together, they can always fuse in general.
The mutual fusion strengths $F_{\gamma_{i}:\gamma_{j}'}$ are summarized
in Table\ \ref{TableII} and depend sinusoidally on the pairing phase
difference $2\delta\Phi$ and the chemical potentials $\mu$ and $\mu'$
of the two islands.

\begin{table}[h]
\begin{tabular}{|r@{\extracolsep{0pt}.}l||r@{\extracolsep{0pt}.}l|r@{\extracolsep{0pt}.}l|r@{\extracolsep{0pt}.}l|r@{\extracolsep{0pt}.}l|}
\hline
\multicolumn{2}{|c||}{\begin{turn}{90}
\end{turn}} & \multicolumn{2}{c|}{$\gamma_{1}$} & \multicolumn{2}{c|}{$\gamma_{2}$} & \multicolumn{2}{c|}{$\gamma_{3}$} & \multicolumn{2}{c|}{$\gamma_{4}$}\tabularnewline
\hline
\multicolumn{2}{|c||}{\,$\gamma_{1}'$\ } & \multicolumn{2}{c|}{$\sin\vartheta_{-}\sin\delta\Phi$} & \multicolumn{2}{c|}{$\cos\vartheta_{+}\cos\delta\Phi$} & \multicolumn{2}{c|}{$\cos\vartheta_{-}\sin\delta\Phi$} & \multicolumn{2}{c|}{$\sin\vartheta_{+}\cos\delta\Phi$}\tabularnewline
\hline
\multicolumn{2}{|c||}{$\gamma_{2}'$} & \multicolumn{2}{c|}{$\cos\vartheta_{+}\cos\delta\Phi$} & \multicolumn{2}{c|}{$\sin\vartheta_{-}\sin\delta\Phi$} & \multicolumn{2}{c|}{$\sin\vartheta_{+}\cos\delta\Phi$} & \multicolumn{2}{c|}{$\cos\vartheta_{-}\sin\delta\Phi$}\tabularnewline
\hline
\multicolumn{2}{|c||}{$\gamma_{3}'$} & \multicolumn{2}{c|}{$\cos\vartheta_{-}\sin\delta\Phi$} & \multicolumn{2}{c|}{$\sin\vartheta_{+}\cos\delta\Phi$} & \multicolumn{2}{c|}{$\sin\vartheta_{-}\sin\delta\Phi$} & \multicolumn{2}{c|}{$\cos\vartheta_{+}\cos\delta\Phi$}\tabularnewline
\hline
\multicolumn{2}{|c||}{$\gamma_{4}'$} & \multicolumn{2}{c|}{$\sin\vartheta_{+}\cos\delta\Phi$} & \multicolumn{2}{c|}{$\cos\vartheta_{-}\sin\delta\Phi$} & \multicolumn{2}{c|}{$\cos\vartheta_{+}\cos\delta\Phi$} & \multicolumn{2}{c|}{$\sin\vartheta_{-}\sin\delta\Phi$}\tabularnewline
\hline
\end{tabular}

\label{tab:fusion-rule}

\caption{Fusion strength $F_{\gamma_{i}:\gamma_{i'}}$ of MBSs $\{\gamma_{i}\}$ and $\{\gamma_{i}'\}$ belonging to two SOTS islands. The table displays the dependence of $F_{\gamma_{i}:\gamma_{i'}}$ on $\mu$ and $\mu'$ and on $\delta\Phi$. We define $\vartheta_{\pm}=(\vartheta\pm\vartheta')/2$, $\vartheta=\text{arctan}(\mu/\Delta_{0})$ and $\vartheta'=\text{arctan}(\mu'/\Delta_{0})$. Results for the fusion of MBSs belonging to the same island can be obtained by taking $\gamma_{i}'=\gamma_{i}$, $\delta\Phi=0$ and $\mu'=\mu$.}

\label{TableII}
\end{table}

\begin{figure}[t]
\includegraphics[width=0.98\columnwidth]{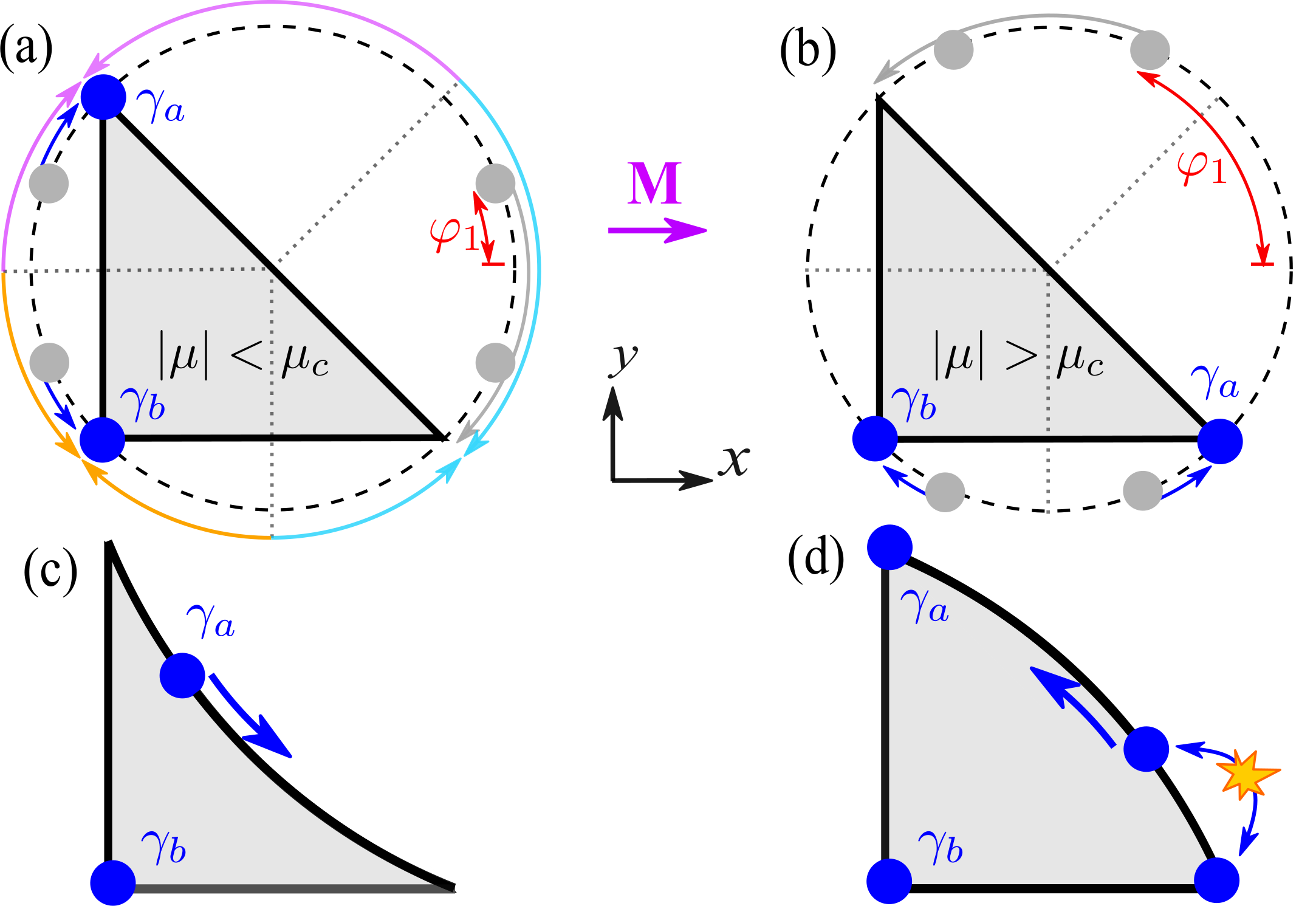}

\caption{Positions of two MBSs (blue dots) in an IRTI for (a) $|\mu|<\mu_{c}$ and (b) $|\mu|>\mu_{c}$, respectively. The gray dots denote the four
MBSs on a disk boundary (dashed curve). The magnetic order (${\bf M}$) is fixed in $x$ direction. By increasing $\mu$ from $0<\mu<\mu_{c}$ to $\mu>\mu_{c}$, the angle $\varphi_{1}$ (in red) increases and one MBS is moved from one sharp-angle vertex to the other one. Schematics of IRTIs with small concavity (c) or convexity (d) on the diagonals.}

\label{fig:triangle}
\end{figure}

\textit{Manipulation of MBSs in IRTIs}.\textemdash In order to obtain a scalable platform hosting any even number of MBSs which are manipulable by purely electrical means, it is essential to go beyond the simple disk geometry presented so far. Particularly, we focus on IRTIs, the short sides of which are orientated in $x$ and $y$ directions, as depicted in Fig.\ \ref{fig:triangle}. To develop some intuition about the appearance of MBSs in the IRTIs, one can relate the latter to the disk geometry in the following way: the dotted lines normal to the triangle sides define three arcs of the disk boundary (dashed curves); all the points belonging to the same arc reduce to the corresponding vertex of the triangle (colored arrows); conversely, each side of the triangle reduces to a single point on the disk. Out of the four MBSs (gray dots) hosted by the disk, two of them must locate on the same arc meaning that, in the triangle, they fuse on the same vertex. By contrast, the two remaining MBSs locate on different arcs and thus stay robustly as zero-energy corner states (blue dots) in the IRTI. Which vertices host the MBSs crucially depends on the angles $\varphi_{i}$ ($\varphi_{1}$ is depicted in red) and, therefore, on the value of the chemical potential $\mu$.

For $|\mu|<\mu_{c}\equiv(M_{0}^{2}/2-\Delta_{0}^{2})^{1/2}$, the four MBSs on the disk are sketched in Fig.\ \ref{fig:triangle}(a). For $|\mu|>\mu_{c}$, the MBSs are located as shown in Fig.\ \ref{fig:triangle}(b). By slowly tuning $\mu$ across $\mu_{c}$, say from $\mu_{\text{d}}(<\mu_{c})$ to $\mu_{\text{u}}(>\mu_{c})$, we can thus adiabatically move one MBS between two sharp-angle vortices while the other one stays fixed at the right-angle vertex. We observe that a finite $\mu_{c}$ requires $M_{0}>\sqrt{2}\Delta_{0}$. When $\mu$ is close to $\mu_{c}$, the localization length of the movable MBS along the diagonal is approximately proportional to $A\Delta_{0}/(\mu_{c}|\mu-\mu_{c}|)$. Therefore, larger islands pose weaker constraints on the difference $\mu_{\text{u}}-\mu_{\text{d}}$. The possibility to move MBSs between two vertices is confirmed numerically \citep{SuppInf}. These results apply to any IRTIs with the short sides in $x$ and $y$ directions.

To get more insights into the fundamental role played by the SOTS geometry and to make further use of it, we consider a small bending on the diagonal. Interestingly, we find that small concavity on the diagonal allows us to smoothly move the MBS along the diagonal [Fig.\ \ref{fig:triangle}(c)]. It also helps to enhance the excitation gap that protects the MBSs since the diagonal becomes fully gapped everywhere except for one point in space even at $\mu_c$ \cite{SuppInf}. By contrast, small convexity tends to nucleate an extra Majorana pair and thus momentarily increases the ground-state degeneracy [Fig.\ \ref{fig:triangle}(d)].

\begin{figure}[t]
\includegraphics[width=1\columnwidth]{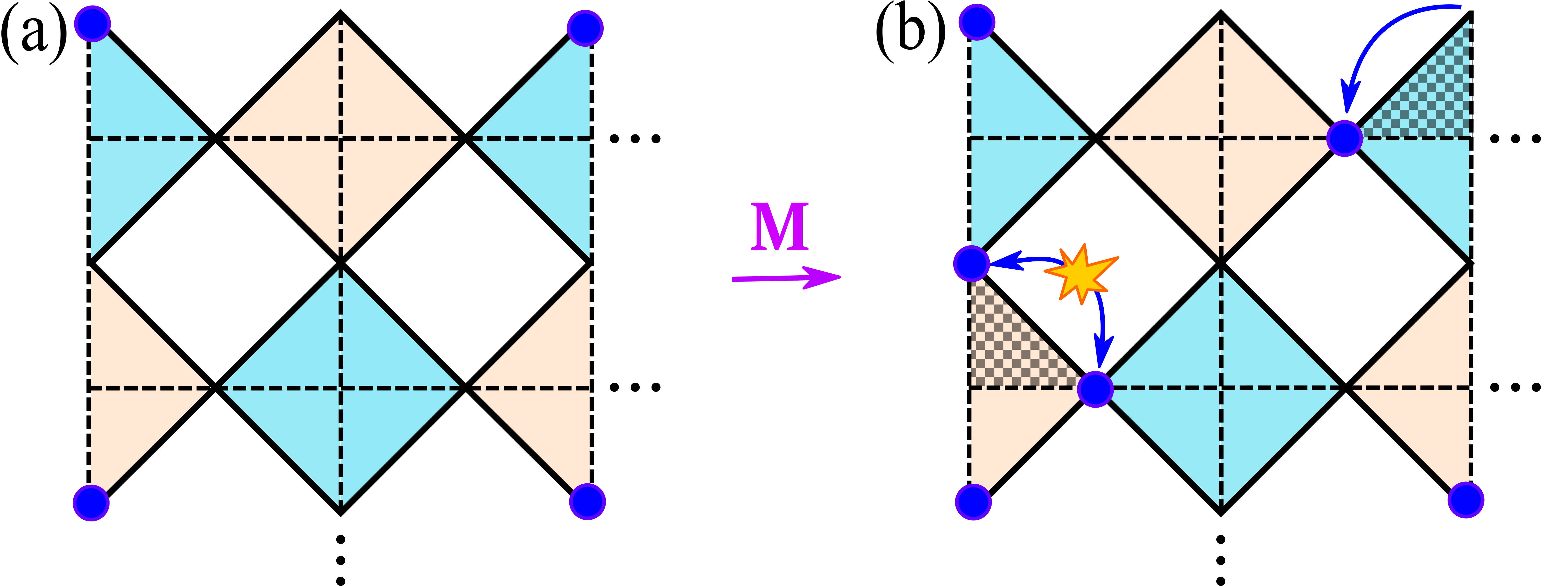}

\caption{Networks of connected IRTIs. The cyan and yellow colors distinguish between two pairing phases on the islands. The white regions are vacuum or trivial insulator. The dashed lines mark the boundaries between the IRTIs.  In (a), all $\mu_j=\mu_{\text{d}}$ and the network hosts four MBSs indicated by the blue dots. In (b), the chemical potentials of triangles marked by the shadow pattern have been tuned to $\mu_{\text{u}}$, resulting in the movement of the top-right MBS and in the nucleation of two additional MBSs.}

\label{fig:scalable}
\end{figure}

\textit{Building networks of MBSs.}\textemdash By properly connecting several IRTIs, networks of diagonals can be defined, for instance, as sketched in Fig.\ \ref{fig:scalable} (more examples are given in the Supplemental Material \citep{SuppInf}). When two or more vertices get in contact, there is a finite overlap between the wavefunctions of different MBSs, which fuse according to the inter- and intra-island fusion strengths summarized in Table \ref{TableII}. The latter clearly depends on the chemical potential and the superconducting phases $\Phi_{j}$ of adjacent IRTIs. For concreteness, in the following, we focus on the configuration illustrated in Fig.\ \ref{fig:scalable}, where we apply $\Phi_{j}=0$ for the cyan triangles and $\Phi_{j}=\Phi_{0}\neq p\pi$ (with $p\in\mathbb{Z}$) for the yellow ones. As a result, we observe that every time an even number of MBSs approach the same point, they completely fuse. Conversely, when an odd number of MBSs approach the same point, a single MBS is left at zero energy.

By tuning the chemical potentials of individual IRTIs across $\mu_{c}$, it is therefore possible to either nucleate, fuse, or move MBSs on
the network. Two clarifying examples are illustrated in Fig.\ \ref{fig:scalable}. In Fig.\ \ref{fig:scalable}(a), all chemical potentials are set to $\mu_{\text{d}}$$<$$\mu_{c}$, resulting in the presence of four MBSs. In Fig.\ \ref{fig:scalable}(b), the chemical potentials of two IRTIs (highlighted by shadow pattern) have been tuned to $\mu_{\text{u}}$$>$$\mu_{c}$. Consequently, the top-right MBS is moved while a new pair of MBSs has been nucleated in the left-bottom of the network.

It is important to stress that the Majorana manipulation does not rely on fine-tuning of parameters. The proposed setup can therefore be easily scaled up, just by adding more IRTIs, in order to accommodate an arbitrary number of MBSs. Since each MBS is exponentially localized on a specific node of the network, the lifting of the ground-state degeneracy is exponentially small in the size of each island.

\begin{figure}[b]
\includegraphics[width=1\columnwidth]{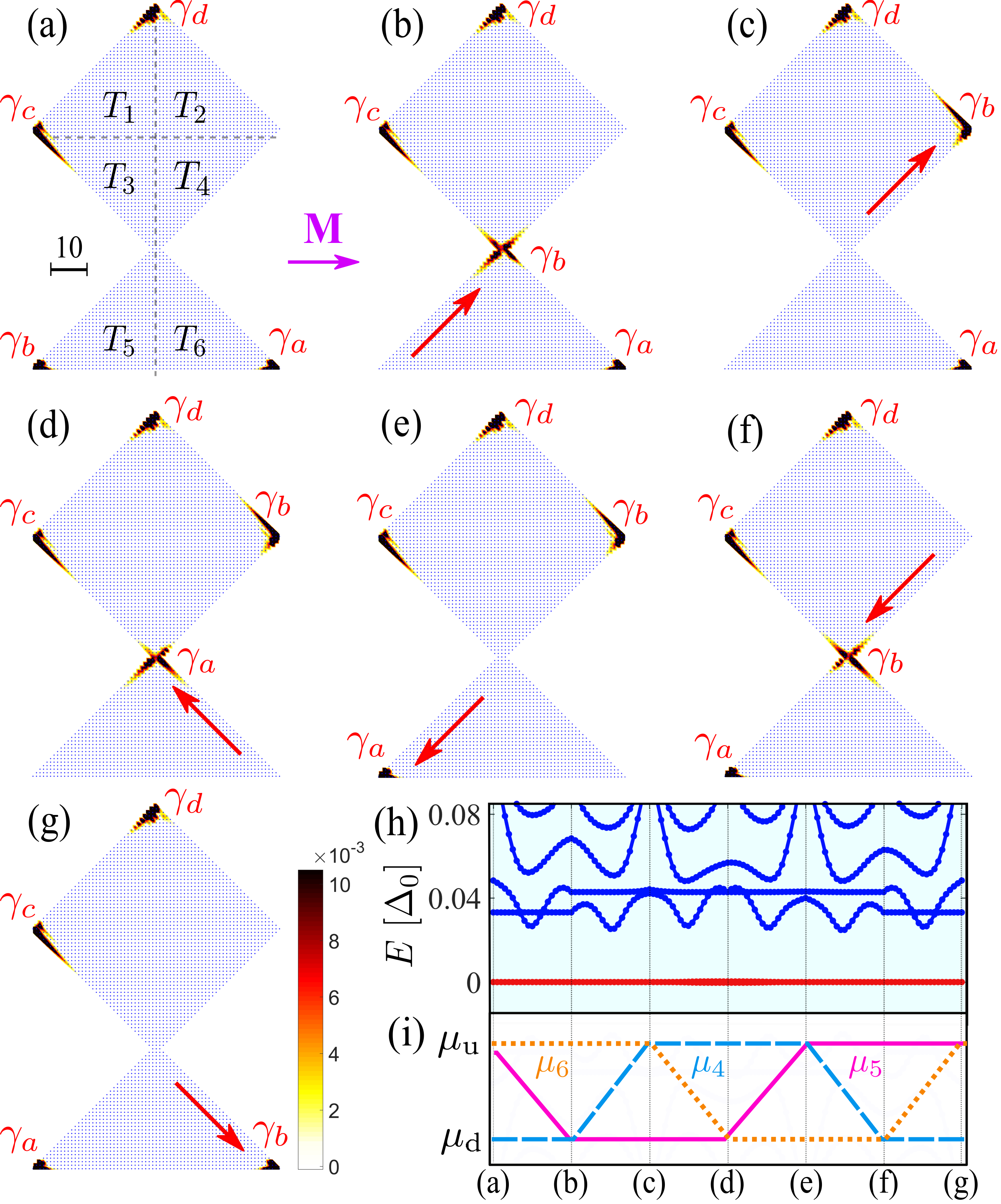}

\caption{Numerical simulation of braiding $\gamma_{a}$ and $\gamma_{b}$.  (a)-(g) Seven subsequent snapshots show the positions of the four
MBSs (black localized densities). During the protocol, $\mu_{4}$, $\mu_{5}$ and $\mu_{6}$ are varied in time, according to (i), while $\mu_{1}$$=$$\mu_{\text{u}}$ and $\mu_{2}$$=$$\mu_{3}$$=$$\mu_{\text{d}}$ are fixed. (h) The energy spectrum of the system during the process. It is symmetric with respect to zero energy. The parameters are $\mu_{\text{u}}=0.15m_{0}$, $\mu_{\text{d}}=0.05m_{0}$, $M_{0}=0.4m_{0}$, $\Delta_{0}=0.25m_{0}$, and $A=m=0.5m_{0}$,  the short-side length of the
IRTIs is $L=35a$.}

\label{fig:braiding}
\end{figure}

\textit{Braiding a Majorana qubit.}\textemdash To illustrate the capabilities of our networks, we now show how to braid a couple of MBSs, thus implementing a phase gate on a Majorana qubit. The latter consists of four MBSs, which can be hosted by the six-island structure depicted in Fig.\ \ref{fig:braiding}. We label the IRTIs by $T_{j}$ (with $j\in\{1,\cdot\cdot\cdot,6$\}) and the corresponding chemical potentials and superconducting phases by $\mu_{j}$ and $\Phi_{j}$, respectively. For the numerical simulation illustrated in Fig.\ \ref{fig:braiding}, we considered $\Phi_{5}=\Phi_{6}=\pi/2$ and $\Phi_{j}=0$ otherwise.

The initial configuration, Fig.\ \ref{fig:braiding}(a), features $\mu_{j}=\mu_{\text{u}}$ for $j\in\{1,5,6\}$ and $\mu_{j}=\mu_{\text{d}}$
otherwise. We can observe four MBSs which are indicated by the black localized densities and labeled by $\gamma_{a}$, $\gamma_{b}$,
$\gamma_{c}$ and $\gamma_{d}$. In order to braid $\gamma_{a}$ and $\gamma_{b}$, the chemical potentials $\mu_{4}$, $\mu_{5}$ and $\mu_{6}$ must be adiabatically tuned in time, according to Fig.\ \ref{fig:braiding}(i). This results in the motion of $\gamma_{a}$ and $\gamma_{b}$ along the diagonals of $T_{4}$, $T_{5}$ and $T_{6}$, as shown in Figs.\ \ref{fig:braiding} (a)-(g). At the end of the protocol, while the system has the same parameters as in the initial state, the positions of $\gamma_{a}$ and $\gamma_{b}$ are exchanged. Importantly, during the whole process, the four MBSs stay robustly at zero energy [red bands in Fig.\ \ref{fig:braiding}(h)]. They are always separated from excited states (blue bands) by an energy gap. Similar procedures apply to exchange other MBS pairs \citep{SuppInf}.

\begin{figure}[bh]
\includegraphics[width=1\columnwidth]{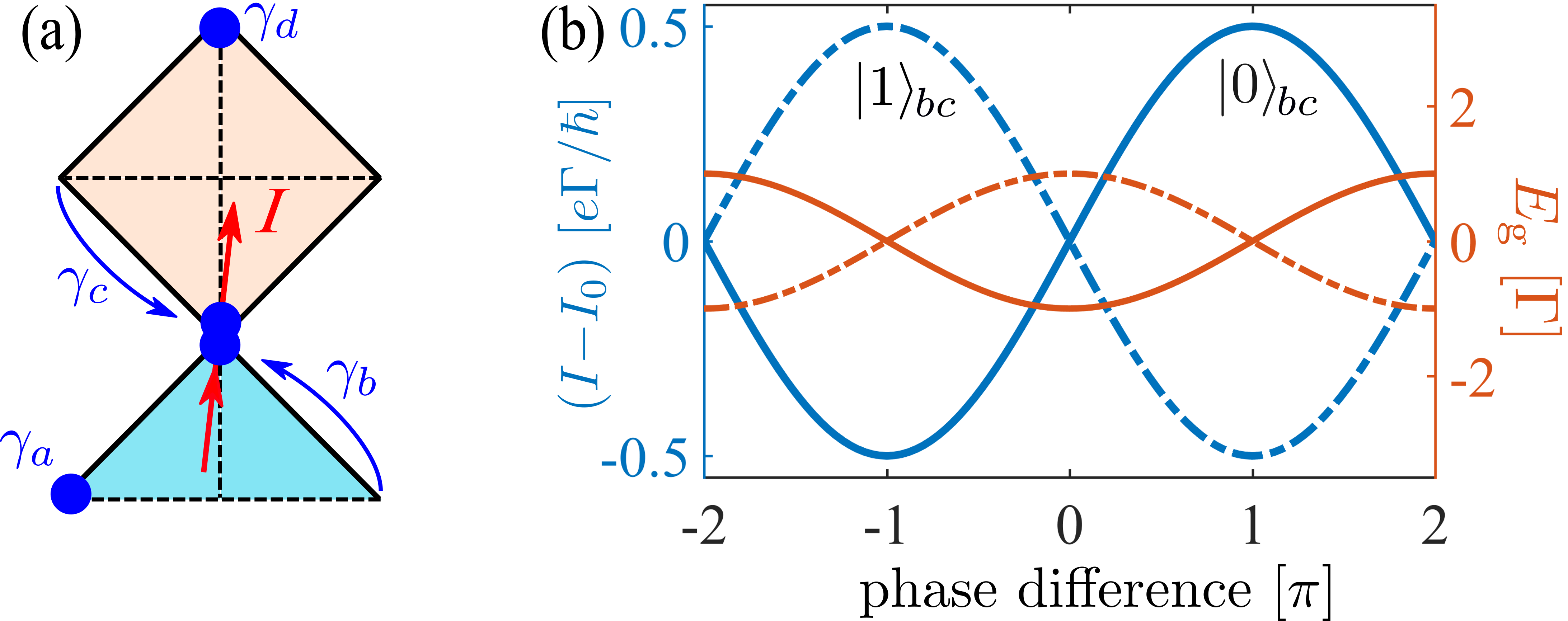}

\caption{(a) Detection of the parity $P_{bc}$ of $\gamma_{b}$ and $\gamma_{c}$
by measuring the Josephson current. (b) Ground-state energy $E_{g}$
(orange curves) of the two coupled MBSs, $\gamma_{b}$ and $\gamma_{c}$,
and the Josephson current (blue curves) across the junction as functions
of the pairing phase difference. Solid and broken curves correspond
to the two parity states, $P_{bc}|0\rangle_{bc}=|0\rangle_{bc}$ and
$P_{bc}|1_{bc}\rangle=-|1_{bc}\rangle$.}

\label{fig:measure-braiding}
\end{figure}

Because of the non-Abelian nature of MBSs, the braiding of $\gamma_{a}$ and $\gamma_{b}$ results in a nontrivial unitary operation $U_{ab}=\exp(\pi\gamma_{a}\gamma_{b}/4)$ on the Majorana qubit \citep{Ivanov01PRL}. It corresponds to a quantum gate that implements a $\pi/2$ rotation on the Bloch sphere. This can be experimentally confirmed by measuring the parity of two different couples of MBSs, $P_{bc}=i\gamma_{b}\gamma_{c}$ and $P_{ac}=i\gamma_{a}\gamma_{c}$. The former one can be used to initialize the qubit, say in the eigenstate of $P_{bc}|0\rangle=|0\rangle$. Then, the braiding rotates the initial state to $U_{ab}|0\rangle$ which is an eigenstate of $P_{ac}U_{ab}|0\rangle=U_{ab}|0\rangle$. The validity of this result can be straightforwardly verified by measuring $P_{ac}$.

Remarkably, our all-in-one setup allows for initialization, braiding, and readout. Indeed, because of the possibility to move and fuse arbitrary
couples of MBSs on the network, we can measure a generic parity operator $P_{\alpha\beta}$. For concreteness, we describe the measurement
of $P_{bc}$ in the six-island architecture. In this case, one must fuse $\gamma_{b}$ and $\gamma_{c}$ by moving them
in the region which defines a Josephson junction between islands with different pairing phases {[}Fig.\ \ref{fig:measure-braiding}(a){]}.
The effective Hamiltonian which describes the coupling between the two MBSs reads $\mathcal{H}_{bc}=\Gamma\cos(\delta\Phi)P_{bc}$, where
$2\delta\Phi$ is the pairing phase difference and $\Gamma$ is the coupling strength that depends on the chemical potentials and
wavefunction overlap. The two eigenenergies are therefore $E_{g}=\pm\Gamma\cos(\delta\Phi)$ [orange curves in Fig.\ \ref{fig:measure-braiding}(b)]. At zero temperature, the Josephson current across the junction is $I=I_{0}\mp e\Gamma\sin(\delta\Phi)/2\hbar=I_{0}\mp I_{\text{mbs}}$, where $I_{\text{mbs}}$ and $I_{0}$ are the contributions from the MBSs and ordinary fermions, respectively \citep{Fu08PRL}. As long as $2\delta\Phi\neq0$, by probing $I$ flowing between the islands one can therefore measure $P_{bc}$ [Fig.\ \ref{fig:measure-braiding}(b)]. In principle, other measurement schemes based on quantum dots are also possible \citep{Flensberg11PRL,SuppInf}.

\textit{Experimental feasibility and summary}.\textemdash  Remarkably, $\text{FeTe}{}_{1-x}\text{Se}_{x}$ monolayers have been shown to possess a band inversion at the $\Gamma$ point \citep{XXWu16PRB,ShiX17SciB,PengXL19PRB} and intrinsic high-temperature superconductivity \citep{LiFS15PRB,NoteN}. The magnetic order may
be induced by putting (anti)ferromagnets, e.g., FeSe or FeTe layers \citep{FJMa09PRL,WBao09PRL,Manna17NC}, on top of $\text{FeTe}{}_{1-x}\text{Se}_{x}$ monolayers or by applying in-plane magnetic fields. We note that the sustenance of superconductivity under strong in-plane magnetic fields in this material has been reported experimentally \citep{Salamon16SRep}. Interestingly, FeSe monolayers  coupling to substrates may have all the desired ingredients for realizing SOTSs (namely, band inversions at the $M$ points, superconductivity \citep{NHao14PRX,NHao19NSR,Wang16NatM} and AFM order \citep{SLHe13nature}) intrinsically within one material. Quantum spin Hall insulators, such as monolayers of 1T'-WTe$_2$ \cite{XFQian14science,SFWu18science,Fei17Nphys,SJTang17nphys, Sajadi18Science,Fatem18science,lupke20Nphys}, inverted Hg(Cd)Te and InAs/GaSb quantum wells \citep{Bernevig06science,Konig07science,CXLiu08PRLb, Knez11prl,Hart14Nphys,Hart17nphys,Ren19Nature}, in proximity to conventional superconductors could offer another candidate system. Notably, electric gating on superconducting 1T'-WTe$_2$ monolayers has already been demonstrated \cite{Sajadi18Science,Fatem18science}.

In general, the control of local chemical potentials on the islands
might be a challenging task. However, it is by no means
necessary to fine tune the chemical potentials to specific values of
$\mu_{\text{u}}$ and $\mu_{\text{d}}$. The only requirements are
(i) the possibility to tune $\mu$ across its critical value, i.e.,
$\mu_{\text{d}}$$<$$\mu_{c}$$<$$\mu_{\text{u}}$ and (ii)
that, at $\mu_{\text{u}}$ and $\mu_{\text{d}}$, the MBSs are
well localized at the vertices of IRTIs. Importantly, we numerically
prove that inhomogeneities of chemical potential within each IRTI
are not detrimental to our proposal \citep{SuppInf}. Finally, we
remark that field effects on (superconducting) thin films have proven
to be a valid alternative to conventional chemical doping in order
to tune the carrier density \citep{Goldman14ARMR,Hanzawa16PNAS,Haenisch2019},
suggesting the feasibility of controlling local chemical potentials
with external gates.

An important issue, when it comes to Majorana-based quantum computation,
is represented by quasiparticle poisoning (QP) \citep{Goldstein11PRB,Budich12PRB,Rainis12PRB},
causing detrimental flips in the total fermion parity of individual
qubits. In this respect, the large superconducting gap of $\text{FeTe}{}_{1-x}\text{Se}_{x}$ monolayers (up to 16.5 meV \citep{LiFS15PRB}) represents a prime
advantage: (i) It is likely to decrease the QP rate.
(ii) It allows for faster adiabatic qubit operations.
Moreover, it might be possible to implement quasiparticle filters
which have proven, at least for quantum wires, to increase the characteristic
QP time up to $(1/200)$s \citep{Menard19PRB}.

In summary, we have proposed a feasible way to realize networks of
SOTSs which can accommodate any even number of topologically protected
MBSs. The MBSs can be generated, moved and fused by all-electrical
means. Our proposal allows to define a qubit, braid the corresponding MBSs,
and measure the nontrivial outcome of this operation.

\begin{acknowledgments}
We thank Sang-Jun Choi, Ning Hao, Tobias
Kiessling, and Wenbin Rui for valuable discussion. This work was supported
by the DFG (SPP1666 and SFB1170 ``ToCoTronics''), the W\"urzburg-Dresden
Cluster of Excellence ct.qmat, EXC2147, project-id 390858490, and the
Elitenetzwerk Bayern Graduate School on ``Topological Insulators''.

{\label{equalCon}S.B.Z. and A.C. contributed equally to this work.}
\end{acknowledgments}

%\bibliographystyle{apsrev4-1}
%\bibliography{ReferenceSB}


\onecolumngrid
\appendix
\clearpage
\newpage
\renewcommand{\thepage}{S\arabic{page}}
\renewcommand{\thefigure}{S\arabic{figure}}

\begin{center}
\bf{\large Supplemental Material}
\end{center}

In this Supplemental Material, we derive the low-energy effective
boundary Hamiltonian and the wavefunctions of Majorana bound states
(MBSs) of the second-order topological superconductors (SOSTs) in
Sec.\ \ref{sec:Derivations-of-effective}. In Sec.\ \ref{sec:Numerical-simulations-of},
w\textcolor{black}{e analyze the motion and the localization properties
of MBSs on a single triangle island when varying the chemical potential,
the excitation energy gap during the Majorana motion, the influence
of adding small bending to the triangle diagonals, and provide numerical
simulations of MBS pairs on the network.} We discuss the measurement
of Majorana qubits via quantum dots in Sec.\ \ref{sec:Measurement-of-Majorana}.
We numerically study the robustness of our results with respect to
moderate inhomogeneities of chemical potential within individual triangle
islands in Sec.\ \ref{sec:robustness-with-respect}. Finally, in
Sec.\ \ref{sec:Other-setups}, we discuss alternative examples of
Majorana networks. \\

\twocolumngrid

\section{Derivations of the effective boundary Hamiltonian and wavefunctions
of Majorana bound states\label{sec:Derivations-of-effective}}

\subsection{Effective boundary Hamiltonian}

In this subsection, we derive the effective boundary Hamiltonian on
a disk geometry. To do so, we first derive the boundary states in
the absence of magnetic and superconducting order. The low-energy
Hamiltonian without magnetic and superconducting order decouples into
four blocks which are respectively for spin-up, spin-down electrons,
spin-up, and spin-down holes. These four blocks are related by time-reversal
and particle-hole symmetries. In the following, we take the block
for spin-up electrons for illustration. In polar coordinates, $r=\sqrt{x^{2}+y^{2}}$
and $\varphi=\arctan(y/x)$, this block Hamiltonian is given by
\begin{equation}
\begin{aligned}h_{e\uparrow}= & \begin{pmatrix}m(\partial^{2}) & -Ae^{-i\varphi}(i\partial_{r}+r^{-1}\partial_{\varphi})\\
-Ae^{i\varphi}(i\partial_{r}-r^{-1}\partial_{\varphi}) & -m(\partial^{2})
\end{pmatrix},\end{aligned}
\end{equation}
where $m(\partial^{2})=m_{0}+m(\partial_{r}^{2}+r^{-2}\partial_{\varphi}^{2}+r^{-1}\partial_{r})$.
In the disk without magnetic order, the angular momentum $\nu$ is
a good quantum number. Consider large radius $R\gg|m/A|$. We can
assume an ansatz for the boundary-state wavefunction as
\begin{align}
\psi({\bf x}) & =(e^{i\nu\varphi}e^{\lambda r}/\sqrt{r})(\alpha,\beta e^{i\varphi})^{T},\label{eq:trial}
\end{align}
where ${\bf x}\equiv(r,\varphi)$. The $\varphi$ periodicity of the
wavefunction $\psi(r,\varphi)=\psi(r,\varphi+2\pi)$ imposes the constraint
$\nu\in\mathbb{Z}\equiv\{0,\pm1,\pm2,\cdots\}$. Plugging the ansatz\ (\ref{eq:trial})
into the Dirac equation for a given energy $\epsilon$, and solving
the equation, we find four solutions of $\lambda$ as $\pm\lambda_{1/2}$,
where
\begin{eqnarray}
\lambda_{1/2}^{2} & = & (\nu+1/2)^{2}/r^{2}-(2mm_{0}-v^{2})/(2m^{2})\nonumber \\
 &  & \pm(A^{4}-4mm_{0}A^{2}-4m^{2}\epsilon_{\nu}^{2})^{1/2}/(2m^{2}),
\end{eqnarray}
and correspondingly four solutions of $(\alpha,\beta)^{T}$, where
$\epsilon_{\nu}=\epsilon-m(\nu+1/2)/r^{2}$. The boundary states are
localized on the boundary. We thus expand the wavefunctions as
\begin{align}
\Psi_{e,\uparrow}({\bf x}) & =\sum_{j=1,2}C_{\lambda_{j}}e^{i\nu\varphi}\dfrac{e^{\lambda_{j}r}}{\sqrt{r}}\begin{pmatrix}iA[\lambda_{j}+(\nu+1/2)/r]\\
\left(m_{\nu}+m\lambda_{j}^{2}-\epsilon_{\nu}\right)e^{i\varphi}
\end{pmatrix},
\end{align}
where $m_{\nu}=m_{0}-m(\nu+1/2)^{2}/r^{2}$ and $\text{Re}[\lambda_{1/2}(R)]>0$
have been assumed without loss of generality. Imposing open boundary
conditions to this wavefunction
\begin{equation}
\Psi_{e,\uparrow}(r=R,\varphi)=0,\label{eq:Bounday-condition}
\end{equation}
the allowed energy of boundary states can be found explicitly as
\begin{eqnarray}
\epsilon(\nu) & = & -\text{sgn}(m)|A|\nu/R+m\nu/R^{2}.\label{eq:eigen-value}
\end{eqnarray}
Hence, the coefficients $C_{\lambda_{1}}$ and $C_{\lambda_{2}}$
are also found. The resulting wavefunctions can be written as
\begin{eqnarray}
\Psi_{e,\uparrow}({\bf x}) & = & e^{i\nu\varphi}K(r)(\text{sgn}(mA),-ie^{i\varphi})^{T},\label{eq:WF}
\end{eqnarray}
where $K(r)=\mathcal{N}[e^{\lambda_{1}(r-R)}-e^{\lambda_{2}(r-R)}]/\sqrt{r}$,
$\lambda_{1/2}=\left|A/(2m)\right|\pm(A^{2}/(4m^{2})-m_{0}/m+(\nu+1/2)^{2}/R^{2})^{1/2}$
and $\mathcal{N}$ is the normalization factor.

For large $R\gg|m/A|$, we approximate a small segment of the disk
boundary as a straight line. Define an effective coordinate and corresponding
momentum as
\begin{eqnarray}
s\equiv R\varphi,\ \ \ p_{\varphi} & \equiv & \nu/R.
\end{eqnarray}
Then, the dispersion relation (\ref{eq:eigen-value}) becomes
\begin{eqnarray}
E_{e,\uparrow}(p_{\varphi}) & = & -\text{sgn}(m)|A|p_{\varphi}-\mu,\label{eq:eigen-energy}
\end{eqnarray}
and the corresponding wavefunction (\ref{eq:WF})
\begin{equation}
\Psi_{e,\uparrow,p_{\varphi}}({\bf x})=e^{ip_{\varphi}s}K(r)(\text{sgn}(mA),-ie^{i\varphi})^{T},
\end{equation}
where $\lambda_{1,2}=\left|A/(2m)\right|\pm(A^{2}/(4m^{2})-m_{0}/m+p_{\varphi}^{2})^{1/2}$
and $K(r)=\mathcal{N}[e^{\lambda_{1}(r-R)}-e^{\lambda_{2}(r-R)}]/\sqrt{R}$.
In Eq.\ (\ref{eq:eigen-energy}), we have considered the chemical
potential $\mu$. In the full basis of the bulk Hamiltonian, the wavefunction
reads
\begin{align}
\Psi_{e,\uparrow,p_{\varphi}}({\bf x}) & =e^{ip_{\varphi}s}K(r)(\text{sgn}(mA),-ie^{i\varphi},0,0,0,0,0,0)^{T}.
\end{align}

Exploiting time-reversal and particle-hole symmetries, we obtain easily
the results for spin-down electrons, spin-up and spin-down holes as
\begin{eqnarray}
E_{e,\downarrow}(p_{\varphi}) & = & E_{e,\downarrow}(-p_{\varphi}),\nonumber \\
E_{h,\uparrow}(p_{\varphi}) & = & -E_{e,\downarrow}(-p_{\varphi}),\nonumber \\
E_{h,\downarrow}(p_{\varphi}) & = & -E_{e,\uparrow}(-p_{\varphi}),
\end{eqnarray}
and
\begin{eqnarray}
\Psi_{e,\downarrow,p_{\varphi}}({\bf x}) & = & is_{y}\Psi_{e,\uparrow,-p_{\varphi}}^{*}({\bf x}),\nonumber \\
\Psi_{h,\uparrow/\downarrow,p_{\varphi}}({\bf x}) & = & \tau_{x}\Psi_{e,\uparrow/\downarrow,-p_{\varphi}}^{*}({\bf x}).
\end{eqnarray}
These boundary states are helical with velocity $A$: the spin-up
electrons move in one direction and the spin-down electrons move in
the opposite direction.

Using the helical states $\text{(\ensuremath{\Psi_{e,\uparrow},\Psi_{e,\downarrow},\Psi_{h,\uparrow},\Psi_{h,\downarrow}})}$
as basis, we now project the superconducting and magnetic order onto
it. The projection can be performed as
\begin{equation}
\widetilde{\mathcal{H}}_{i,j}=\langle\Psi_{i}|(\Delta_{0}\tau_{y}s_{y}+H_{M})|\Psi_{j}\rangle,
\end{equation}
where the subscript $i,j$ are abbreviations of the indices $(e/h,\uparrow/\downarrow)$.
Assuming $mA>0$ without loss of generality, we obtain the effective
boundary Hamiltonian stated in Eq.\ (2) of the main text.

\subsection{Wavefunctions of Majorana bound states\label{sec:Wavefunction-and-polarization}}

Next, we derive the wavefunctions of MBSs. Assume the wavefunction
for the MBS $\gamma_{i}$ around its localization center $\varphi_{i}$
in the form
\begin{equation}
\Psi_{0}=e^{R\int\xi(\varphi)d\varphi}(c_{1},c_{2},c_{3},c_{4})^{T}.
\end{equation}
The eigen equation at zero energy is then given by
\begin{align}
 & \begin{pmatrix}iA\xi-\mu & iM_{0}\sin\varphi & 0 & -\Delta_{0}\\
-iM_{0}\sin\varphi & -iA\xi-\mu & \Delta_{0} & 0\\
0 & \Delta_{0} & iA\xi+\mu & iM_{0}\sin\varphi\\
-\Delta_{0} & 0 & -iM_{0}\sin\varphi & -iA\xi+\mu
\end{pmatrix}\begin{pmatrix}e^{i\varphi/2}c_{1}\\
e^{-i\varphi/2}c_{2}\\
e^{-i\varphi/2}c_{3}\\
e^{i\varphi/2}c_{4}
\end{pmatrix}\nonumber \\
 & =0.\label{eq:Secular-Eq}
\end{align}
Solving Eq.\ (\ref{eq:Secular-Eq}), we obtain four solutions of
$\xi$ as $\pm\xi_{1/2}$ with
\begin{eqnarray}
\xi_{1/2} & = & \Delta_{0}/A\pm\sqrt{M_{0}^{2}\sin^{2}\varphi-\mu^{2}}/A,\label{eq:localization-length}
\end{eqnarray}
and the corresponding four nontrivial solutions of $\Psi_{0}$.

For illustration, let us consider the MBS $\gamma_{1}$ at
\begin{eqnarray}
\varphi_{1} & = & \arcsin(\bar{\Delta}/M_{0}),\ \ \bar{\Delta}=(\Delta_{0}^{2}+\mu^{2})^{1/2}.
\end{eqnarray}
For $\varphi$ slightly larger than $\varphi_{1}$, we find $\text{Re}(-\xi_{1})<0$
and $\text{Re}(\xi_{2})<0$. For $\varphi$ slightly smaller than
$\varphi_{1}$, we obtain $\text{Re}(\xi_{1})>0$ and $\text{Re}(\xi_{2})>0$.
Thus, the wavefunction of $\gamma_{1}$ around $\varphi_{1}$ can
be expanded as
\begin{equation}
\Psi_{1}=\begin{cases}
\alpha_{>}e^{-\int\xi_{1}Rd\varphi}(i,ie^{i\vartheta}e^{i\varphi_{1}},e^{i\vartheta}e^{i\varphi_{1}},1)^{T}\\
+\beta_{>}e^{\int\xi_{2}Rd\varphi}(-i,-ie^{-i\vartheta}e^{i\varphi_{1}},e^{-i\vartheta}e^{i\varphi_{1}},1)^{T}, & \varphi>\varphi_{1}\\
\alpha_{<}e^{\int\xi_{1}Rd\varphi}(-i,ie^{-i\vartheta}e^{i\varphi_{1}},-e^{-i\vartheta}e^{i\varphi_{1}},1)^{T}\\
+\beta_{<}e^{\int\xi_{2}Rd\varphi}(-i,-ie^{-i\vartheta}e^{i\varphi_{1}},e^{-i\vartheta}e^{i\varphi_{1}},1)^{T}, & \varphi<\varphi_{1}
\end{cases}
\end{equation}
where $e^{i\vartheta}=(\Delta_{0}+i\mu)/\bar{\Delta}$. Considering
the continuity of the wavefunction at $\varphi=\varphi_{1}$, we find
$\alpha_{>}=\alpha_{<}=0$ and $\beta_{>}=\beta_{<}$. Therefore,
$\Psi_{1}$ can be simplified to
\begin{eqnarray}
\Psi_{1} & = & e^{\int\xi_{2}Rd\varphi}(-i,-ie^{-i\vartheta}e^{i\varphi_{1}},e^{-i\vartheta}e^{i\varphi_{1}},1)^{T}.
\end{eqnarray}
In the original basis $(c_{a\uparrow},c_{b\uparrow},c_{a\downarrow},c_{b\downarrow},c_{a\uparrow}^{\dagger},c_{b\uparrow}^{\dagger},c_{a\downarrow}^{\dagger},c_{b\downarrow}^{\dagger})$,
the wavefunction can be written as
\begin{eqnarray}
\Psi_{1} & = & e^{\int\xi_{2}Rd\varphi}K(r)(e^{-i(\varphi_{1}-\vartheta+\pi/2)/2},-e^{i(\varphi_{1}+\vartheta+\pi/2)/2},\nonumber \\
 &  & \qquad\qquad e^{i(\varphi_{1}-\vartheta-\pi/2)/2},e^{-i(\varphi_{1}+\vartheta-\pi/2)/2},\nonumber \\
 &  & \qquad\qquad e^{i(\varphi_{1}-\vartheta+\pi/2)/2},-e^{-i(\varphi_{1}+\vartheta+\pi/2)/2},\nonumber \\
 &  & \qquad\qquad e^{-i(\varphi_{1}-\vartheta-\pi/2)/2},e^{i(\varphi_{1}+\vartheta-\pi/2)/2})^{T},
\end{eqnarray}
up to a phase factor.

Similarly, we can derive the wavefunctions of the other three MBSs
as
\begin{eqnarray}
\Psi_{2} & = & e^{-\int\xi_{2}Rd\varphi}K(r)(-e^{-i(\varphi_{2}+\vartheta+\pi/2)/2},e^{i(\varphi_{2}-\vartheta+\pi/2)/2},\nonumber \\
 &  & \qquad\qquad e^{i(\varphi_{2}+\vartheta-\pi/2)/2},e^{-i(\varphi_{2}-\vartheta-\pi/2)/2},\nonumber \\
 &  & \qquad\qquad-e^{i(\varphi_{2}+\vartheta+\pi/2)/2},e^{-i(\varphi_{2}-\vartheta+\pi/2)/2},\nonumber \\
 &  & \qquad\qquad e^{-i(\varphi_{2}+\vartheta-\pi/2)/2},e^{i(\varphi_{2}-\vartheta-\pi/2)/2})^{T},\nonumber \\
\Psi_{3} & = & e^{\int\xi_{2}Rd\varphi}K(r)(e^{-i(\varphi_{3}-\vartheta-\pi/2)/2},e^{i(\varphi_{3}+\vartheta-\pi/2)/2},\nonumber \\
 &  & \qquad\qquad-e^{i(\varphi_{3}-\vartheta+\pi/2)/2},e^{-i(\varphi_{3}+\vartheta+\pi/2)/2},\nonumber \\
 &  & \qquad\qquad e^{i(\varphi_{3}-\vartheta-\pi/2)/2},e^{-i(\varphi_{3}+\vartheta-\pi/2)/2},\nonumber \\
 &  & \qquad\qquad-e^{-i(\varphi_{3}-\vartheta+\pi/2)/2},e^{i(\varphi_{3}+\vartheta+\pi/2)/2})^{T},\nonumber \\
\Psi_{4} & = & e^{-\int\xi_{2}Rd\varphi}K(r)(e^{-i(\varphi_{4}+\vartheta-\pi/2)/2},e^{i(\varphi_{4}-\vartheta-\pi/2)/2},\nonumber \\
 &  & \qquad\qquad e^{i(\varphi_{4}+\vartheta+\pi/2)/2}-e^{-i(\varphi_{4}-\vartheta+\pi/2)/2},\nonumber \\
 &  & \qquad\qquad e^{i(\varphi_{4}+\vartheta-\pi/2)/2},e^{-i(\varphi_{4}-\vartheta-\pi/2)/2},\nonumber \\
 &  & \qquad\qquad e^{-i(\varphi_{4}+\vartheta+\pi/2)/2}-e^{i(\varphi_{4}-\vartheta+\pi/2)/2})^{T}.
\end{eqnarray}
We can observe from these wavefunctions that $\Psi_{3}=-\mathcal{P}\Psi_{1}$
and $\Psi_{4}=\mathcal{P}\Psi_{2}$, where $\mathcal{P}=\sigma_{z}\mathcal{T}_{\varphi\rightarrow\varphi+\pi}$
is the inversion symmetry operator and $\mathcal{T}_{\varphi\rightarrow\varphi+\pi}=e^{-i\pi\tau_{z}s_{z}\sigma_{z}/2}$
shifts the angle $\varphi$ by $\pi$. This indicates that the modes
$\gamma_{1}$ and $\gamma_{2}$ are inversion partners of $\gamma_{3}$
and $\gamma_{4}$, respectively. Therefore, in the SOTS with inversion
symmetry in the bulk, the scattering between $\gamma_{1}$ and $\gamma_{3}$
(or $\gamma_{2}$ and $\gamma_{4}$) is prohibited, as verified by
$F_{\gamma_{1}:\gamma_{3}}=F_{\gamma_{2}:\gamma_{4}}=0.$

The function $\xi_{2}(\varphi)$ in the wavefunctions determines the
localization length of the MBSs along the boundary. It can equivalently
determine the localization length of the MBSs on a straight boundary
which is normal to the azimuthal direction at $\varphi$. According
to Eq.\ (\ref{eq:localization-length}), the localization length
on this straight boundary is approximately given by
\begin{equation}
l_{decay}\approx A\Delta_{0}/(\mu_{\varphi}|\mu-\mu_{\varphi}|),
\end{equation}
for $\mu$ close to $\mu_{\varphi}=\sqrt{M_{0}^{2}\sin^{2}\varphi-\Delta_{0}^{2}}$.
It diverges when $\mu$ approaches the critical value $\mu_{\varphi}$.

\section{Numerical simulations of moving Majorana bound states\label{sec:Numerical-simulations-of}}

\subsection{MBSs in a single isosceles right triangle island}

\begin{figure}[h]
\includegraphics[width=1\columnwidth]{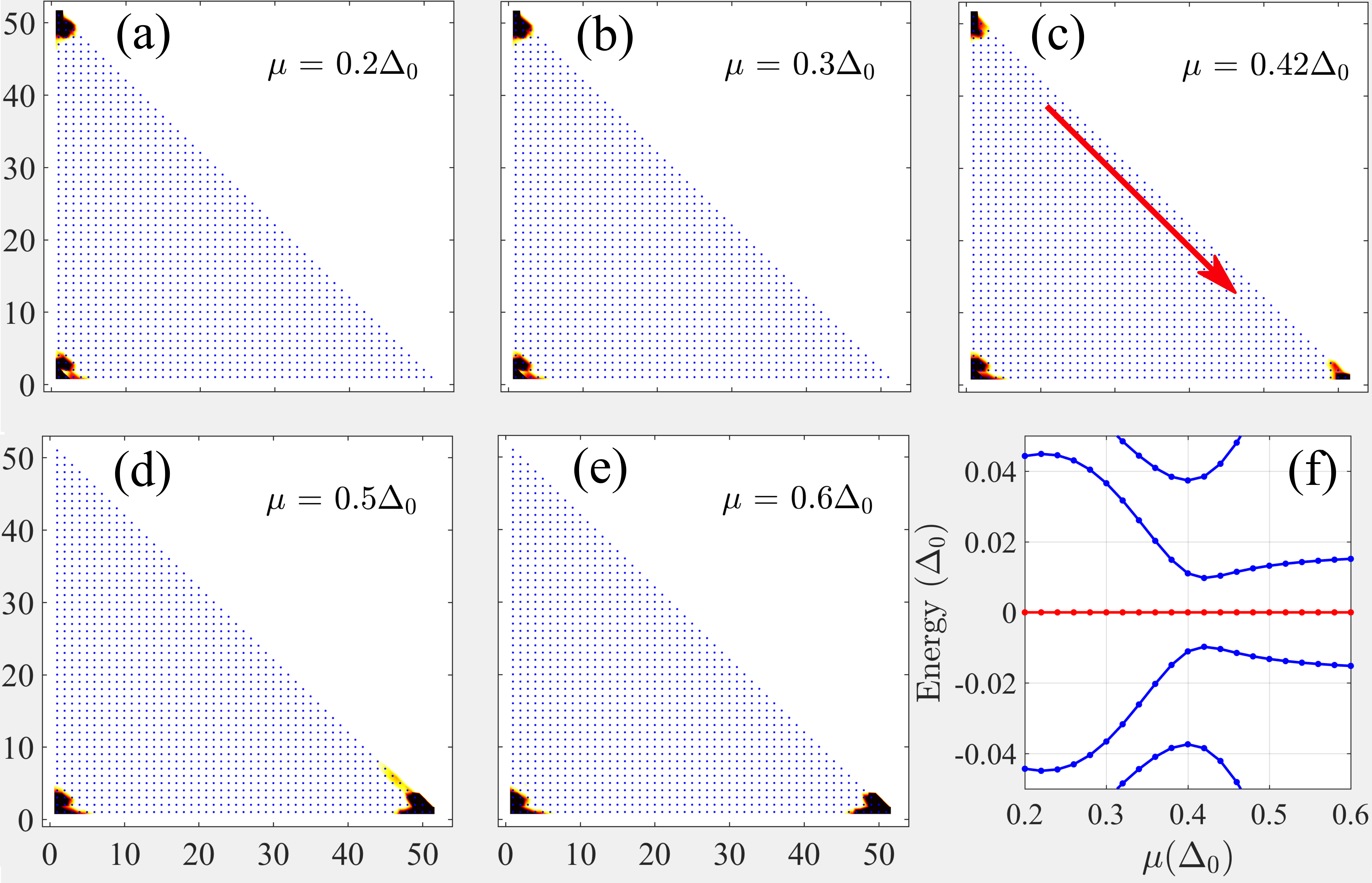}

\caption{\textcolor{black}{(a)-(e) Several subsequent snapshots show the positions
(the black densities) of the MBSs in an IRTI when increasing $\mu$
across $\mu_{c}$. (f) Energy spectrum of this process. At the critical
chemical potential $\mu_{c}$, the MBS on the diagonal is partially
localized at one sharp-angle vertex and partially at the other sharp-angle
vertex. The parameters are $\Delta_{0}=0.25m_{0}$, $M_{0}=0.4m_{0}$,
$A=m=0.5m_{0}=1$, and the short-side length of the IRTI is $L=50a$
with $a$ the lattice constant.}}

\label{fig:straight-diagonal}
\end{figure}

In this subsection, we study numerically the motion of MBSs in a single
isosceles right triangle island (IRTI). We discretize the bulk Hamiltonian
onto a tight-binding lattice (by replacing $k_{i}\rightarrow\sin(k_{i}a)/a$
and $k_{i}^{2}\rightarrow2[1-\cos(k_{i}a)]$, $i\in\{x,y\}$) and
consider the short-side length of the triangle as $L=50a$ for concreteness.
Several subsequent snapshots of the positions of the two MBSs in an
IRTI when increasing the chemical potential are displayed in Fig.\ \ref{fig:straight-diagonal}.
Other parameters are given in the caption. The corresponding animation
that shows the slow movement of MBSs is provided in the supplemental
file ``Triangle-movie.mp4''. In this simulation, we see clearly
that one MBS is moved slowly from one sharp-angle vertex to the other
one along the diagonal while the other MBS is kept untouched at the
right-angle vertex. This behavior is of perfect consistency with our
analytical conclusion.

\begin{figure}[th]
\includegraphics[width=0.92\columnwidth]{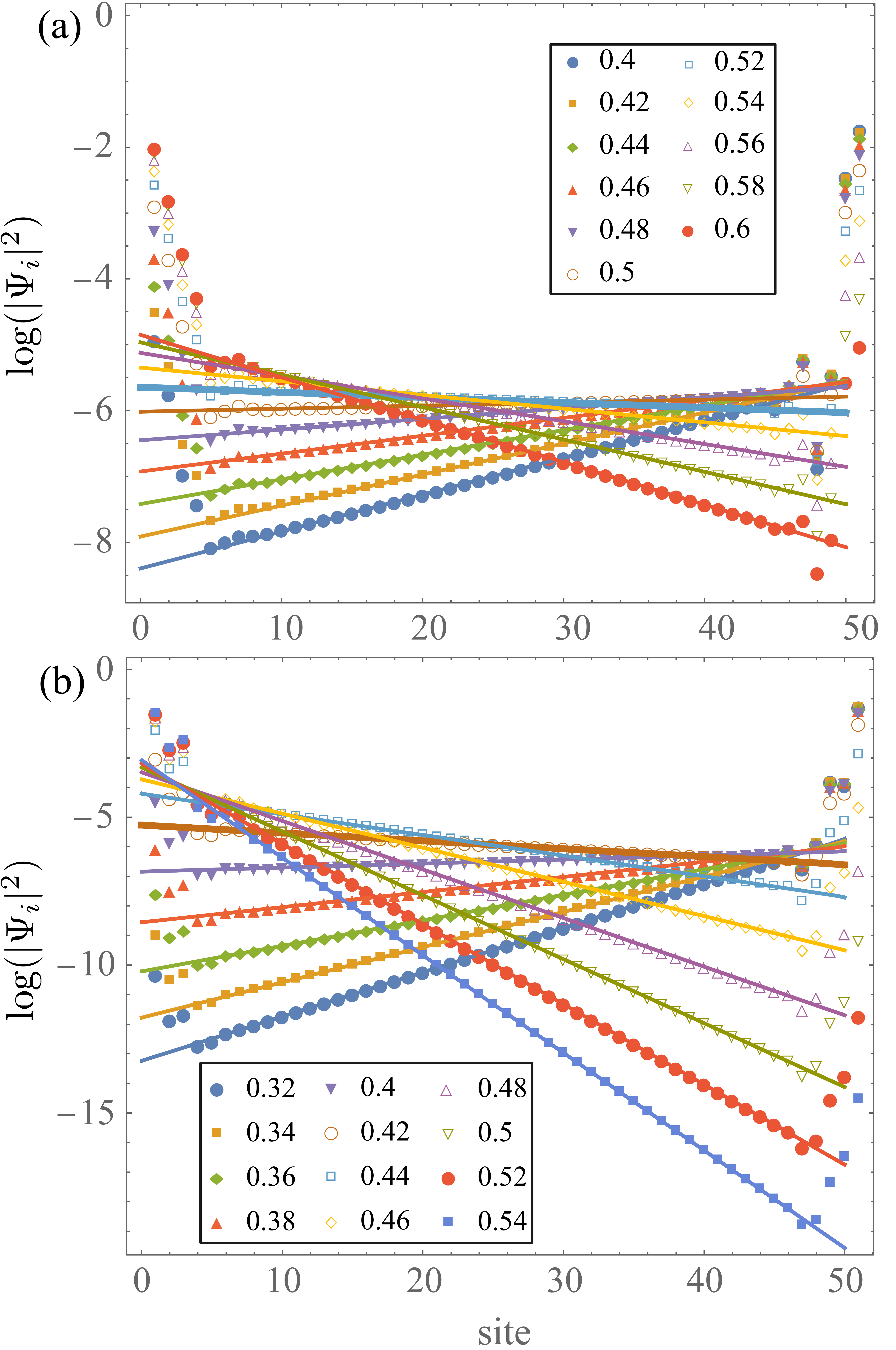}

\caption{\textcolor{black}{Logarithm of the probability density of the movable
Majorana wavefunction along the diagonal of an IRTI for different
chemical potentials (indicated in the legend and in units of $\Delta_{0}$).
The colored dots are numerical data and the lines are linear fittings
on the sites between $10<i<40$. The value of $\mu$ corresponding
to the flat fitting line gives the critical chemical potential $\bar{\mu}_{c}.$
When $\Delta_{0}$ and $M_{0}$ are much smaller than the insulating
gap of the bulk $(\sim m_{0})$, $\bar{\mu}_{c}$ agrees excellently
with the analytical result $\mu_{c}$ ($=0.53\Delta_{0}$), see the
curve of $\mu=0.52\Delta_{0}$ (a). In contrast, when $\Delta_{0}$
and $M_{0}$ become comparable with $m_{0}$, $\bar{\mu}_{c}$ deviates
evidently from $\mu_{c}$, see the curve of $\mu=0.42\Delta_{0}$
in (b). $\Delta_{0}=0.1m_{0}$ for (a) and $\Delta_{0}=0.25m_{0}$
for (b), other parameters are $M_{0}=1.6\Delta_{0}$, $A=m=0.5m_{0}$
and the short-side length of the IRTI is $L=50a$.}}

\label{fig:localization-length}
\end{figure}

Figure\ \ref{fig:localization-length} shows the logarithm of the
probability density {[}$\text{log}(|\Psi_{i}|^{2})${]} of the movable
Majorana wavefunction along the diagonal for different chemical potentials
$\mu$, where $i$ labels the $y$ coordinate of the lattice sites.
The lines are linear fittings on the sites between $10<i<40$. In
the central part of the diagonal, the decay of the wavefunction is
clearly exponential, as expected. By increasing $\mu$, we can observe
that the MBS localizing at the left edge ($i=50)$ moves to localize
at the right one ($i=0$). \textcolor{black}{The localization length
of the MBS wavefunction can be extracted as the inverse of the slopes
of the fitting lines. It diverges at $\bar{\mu}_{c}$ which corresponds
to the critical value of chemical potential. When $\Delta_{0}$ and
$M_{0}$ are small compared to $m_{0}$, $\bar{\mu}_{c}$ agrees excellently
with its analytical value given by $\mu_{c}=\sqrt{M_{0}^{2}/2-\Delta_{0}^{2}}$,
see Fig.\ \ref{fig:localization-length}(a). In contrast, when $\Delta_{0}$
and $M_{0}$ are comparable with $m_{0}$, the critical chemical potential
becomes significantly smaller than $\mu_{c}$, see Fig.\ \ref{fig:localization-length}(b).
This deviation could be attributed to higher-order momentum corrections
in the tight-binding calculation.}

\begin{figure}[th]
\includegraphics[width=0.85\columnwidth]{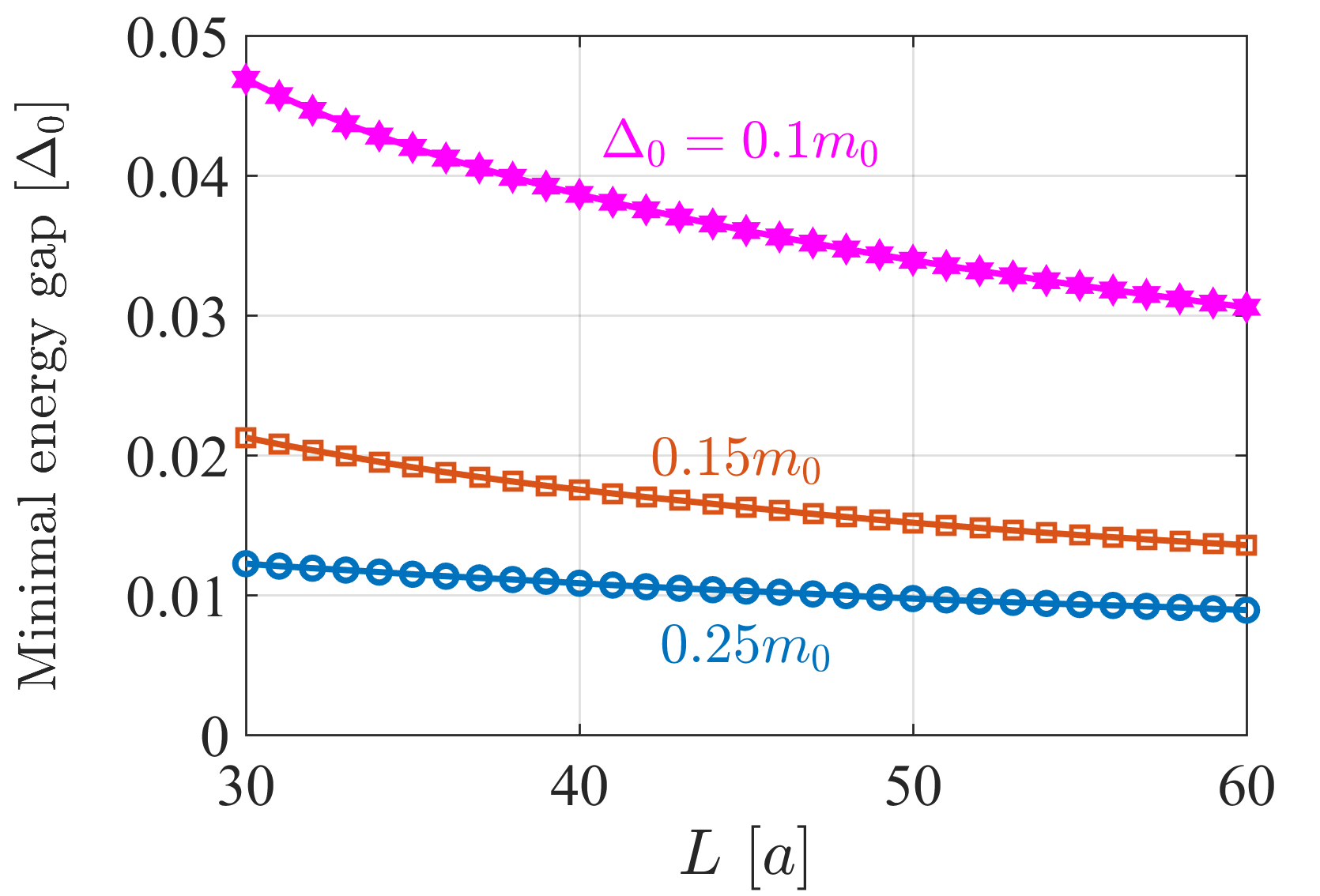}

\caption{\textcolor{black}{The minimal energy gap $\Delta E$ as a function
of the short-side length $L$ of the IRTI. The three curves correspond
to three different pairing potentials, $\Delta_{0}=0.1m_{0}$, $0.15m_{0}$
and $0.25m_{0}$, respectively. $\Delta E$ decreases monotonically
with increasing $L$ but at a rate less pronounced than $1/L$. Other
parameters for this numerical calculation are $M_{0}=1.6\Delta_{0}$
and $A=m=0.5m_{0}$.}}

\label{fig:Gap-Lr}
\end{figure}

\textcolor{black}{At the critical chemical potential $\mu_{c}$, the
energy gap is minimal. This may be related to the fact that at $\mu=\mu_{c}$,
the 1D diagonal edge effectively realizes a topological phase transition.
It becomes gapless if it is infinitely long. A finite length of the
diagonal, however, gives rise to an energy gap $\Delta E$, which
is larger than $\geq A\pi/\sqrt{2}L$, where $L$ is the short-side
length of the IRTI. For inverted Hg(Cd)Te quantum wells, $A>0.3$
eV$\cdot$nm \citep{Bernevig06science} and for FeTe$_{1-x}$Se$_{x}$,
$A>0.2$ eV$\cdot$nm \citep{XXWu16PRB}. Thus, the estimated energy
gap due to the 1D finite-size confinement in these two candidate materials
could be larger than $0.67$ meV (8 K) and $0.44$ meV (5 K) for a
length of $L=1$ $\mu$m, respectively. In Fig.\ \ref{fig:Gap-Lr},
we calculate numerically this minimal energy gap for increasing sizes
of the triangle $L$. Interestingly, it shows an even less pronounced
dependence on $L$ (compared to the $\propto1/L$ dependence).}

\subsection{\textcolor{black}{Effect of bending the diagonal}}

\begin{figure}[th]
\textcolor{red}{\includegraphics[width=0.9\columnwidth]{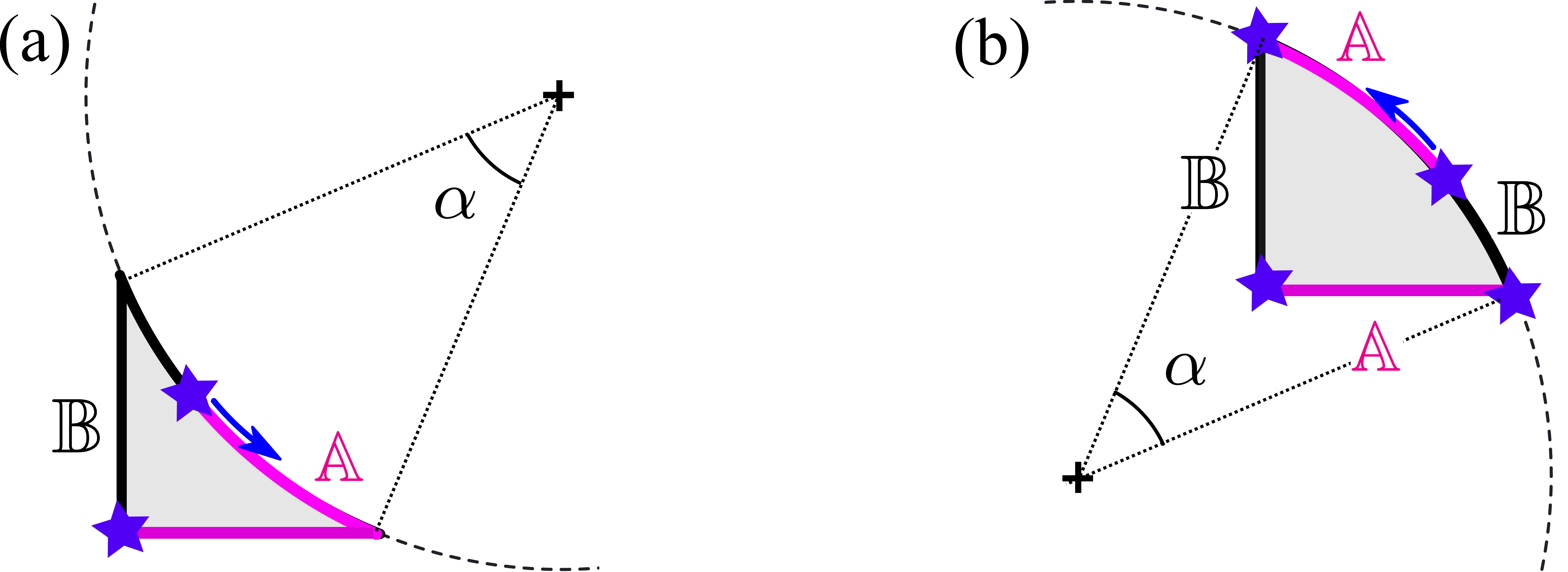}}

\caption{\textcolor{black}{Schematics of IRTIs with small concavity (a) or
convexity (b) bending on the diagonal. The purple and black colors
indicate that the boundary segments belong to classes $\mathbb{A}$
and $\mathbb{B}$, respectively. $\alpha$ denotes the angle of the
diagonal arc. The pentagons represent the MBSs. The MBS on the diagonal
arc can be smoothly moved by tuning $\mu$.}}

\label{fig:bending}
\end{figure}

\textcolor{black}{In this section, we discuss the effect of small
diagonal concavity and convexity bending on the IRTI, as sketched
in Fig.\ \ref{fig:bending}. According to our effective boundary
Hamiltonian, we can classify the boundary into two distinct classes:
$\mathbb{A}$ with $\widetilde{M}>\bar{\Delta}$, and $\mathbb{B}$
with $\widetilde{M}<\bar{\Delta}$, where $\bar{\Delta}=\sqrt{\Delta_{0}^{2}+\mu^{2}}$
and $\widetilde{M}$ is the local effective magnetization. Then, an
MBS forms and only forms at the domain that connects the boundaries
of different classes.}

\begin{figure}[th]
\textcolor{red}{\includegraphics[width=1\columnwidth]{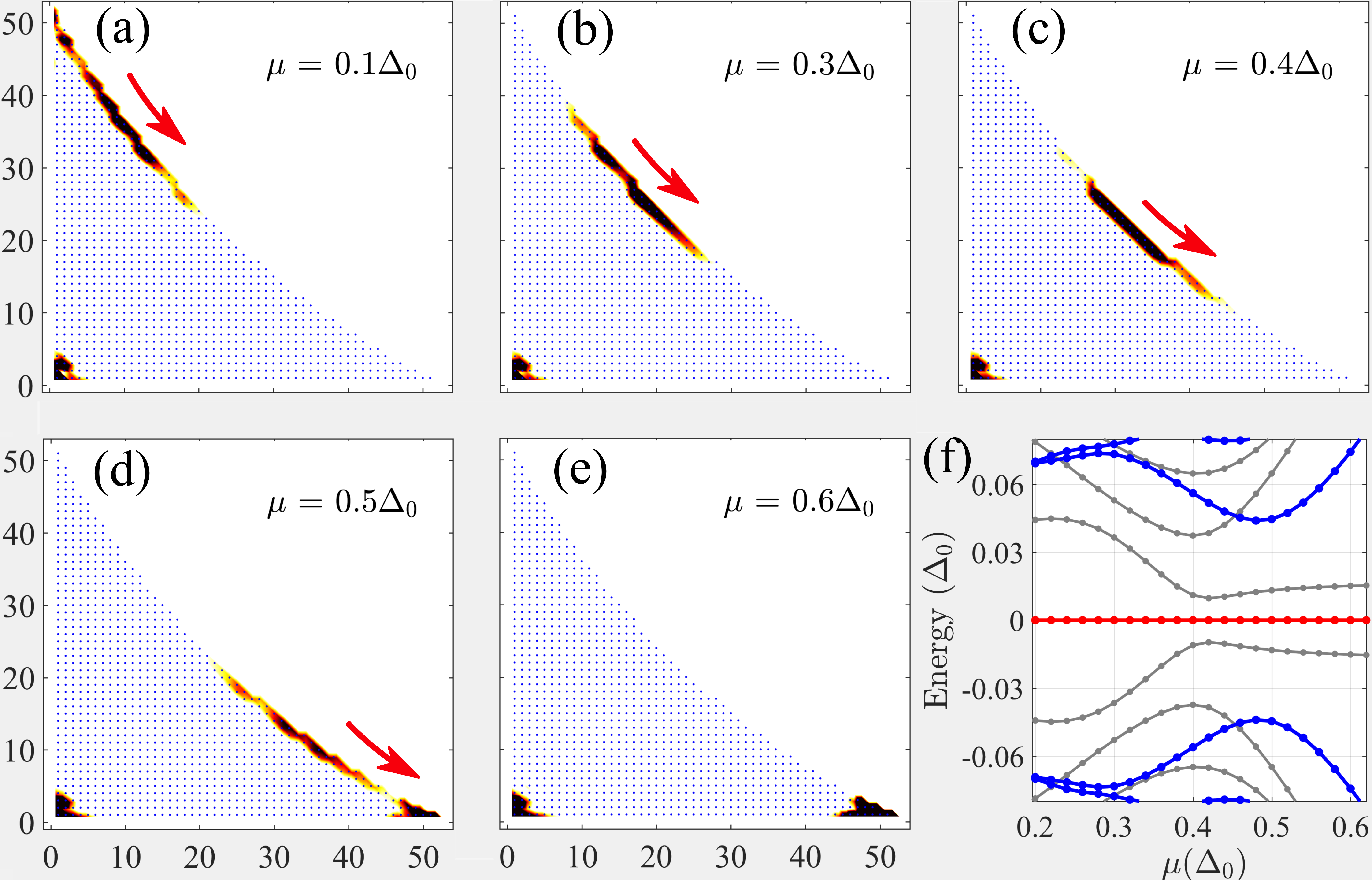}}

\caption{\textcolor{black}{(a)-(e) Several subsequent snapshots show the positions
of the MBSs in an IRTI with a concavity bending of angle $\alpha=0.18\pi$
in the diagonal. The corresponding chemical potentials are given in
the panels. One MBS moves along the diagonal arc from one vertex to
another vertex. (f) Energy spectrum (blue and red curves) of the process.
The gray curves are for an IRTI without bending and are presented
for comparison. The concavity bending in the diagonal significantly
enhances the excitation gap that protects the MBSs in the system.
Other parameters are the same as those in Fig.\ \ref{fig:straight-diagonal}.}}

\label{fig:Concavity}
\end{figure}

\textcolor{black}{In the case of concavity bending, the two short
sides of the IRTI belong to classes $\mathbb{A}$ and $\mathbb{B}$,
respectively. The diagonal arc, in general, is divided into two segments,
one segment belongs to $\mathbb{A}$ and the other segment belongs
to $\mathbb{B}$, as shown in Fig.\ \ref{fig:bending}(a). We have
thus two MBSs, one stays fixed at the right-angle vertex and the other
at the separating point on the diagonal. By tuning the chemical potential,
we can move the separating point on the diagonal. Accordingly, one
MBS moves along the diagonal. Note that the triangle diagonal behaves
like a T-junction. Remarkably, only one single gate is needed for
controlling the MBS, different from the T-junction of semiconducting
nanowires which require many gates in a keyboard form \citep{Alicea11Nphys}.
These results are confirmed numerically in Fig.\ \ref{fig:Concavity}.
It is important to note that the concavity bending on the diagonal
can significantly enhance the energy gap that protects the zero-energy
MBSs, compared to the case of straight diagonal, see Fig.\ \ref{fig:Concavity}(f).
This enhancement can be attributed to the fact that the bent diagonal
becomes gapped everywhere except at the separating point, even at
the critical chemical potential.}

\begin{figure}[th]
\textcolor{red}{\includegraphics[width=1\columnwidth]{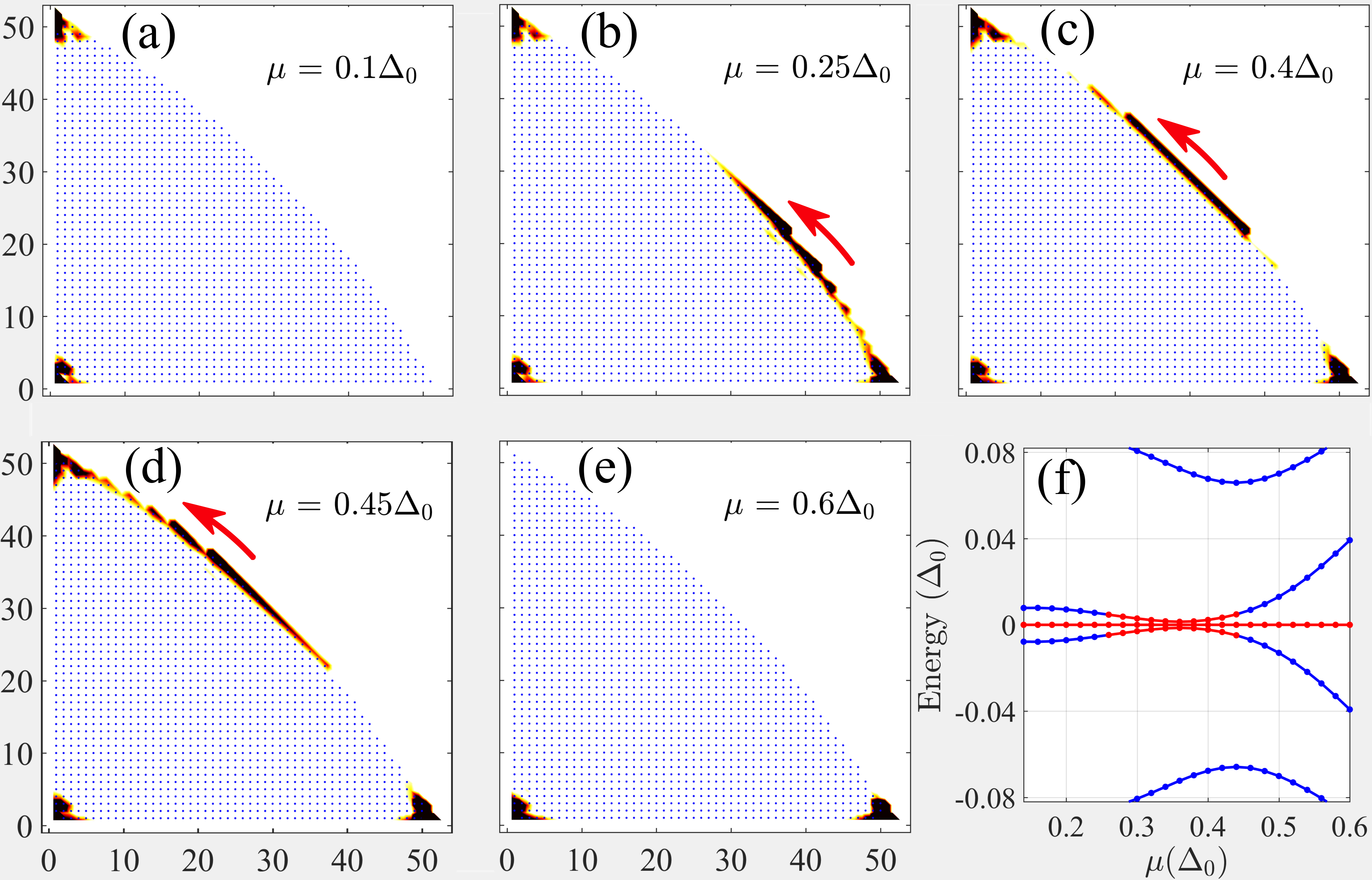}}

\caption{\textcolor{black}{(a)-(e) Several subsequent snapshots show the positions
of the modes with energy lower than $0.005\Delta_{0}$ in an IRTI
with a convexity bending of angle $0.18\pi$ in the diagonal. (f)
Energy spectrum of the process. Increasing $\mu$, an extra pair of
MBSs appear at the empty sharp-angle vertex. One of them moves along
the diagonal arc to the other sharp-angle vertex and annihilates with
the MBS there. Other parameters are the same as those in Fig.\ \ref{fig:straight-diagonal}.}}

\label{fig:Convexity}
\end{figure}

\textcolor{black}{In the case of convexity bending, the diagonal arc
is also divided into two segments but with exchanging their positions,
as compared to the case of concavity bending. Thus, the triangle has
four domains connecting the $\mathbb{A}$ and $\mathbb{B}$ boundaries
and hence hosts four MBSs, see Fig.\ \ref{fig:bending}(b). Three
of the four MBSs stay at the three vertices of the triangle, respectively,
and the other on the diagonal arc. The latter is movable by adjusting
the chemical potential. When it is close to another MBS at the sharp-angle
vertex, they annihilate together. We confirm these behaviors numerically
in Fig.\ \ref{fig:Convexity}.}

\subsection{Numerical simulations of exchanging MBS pairs in the illustrative
network}

\begin{figure}[th]
\includegraphics[width=0.95\columnwidth]{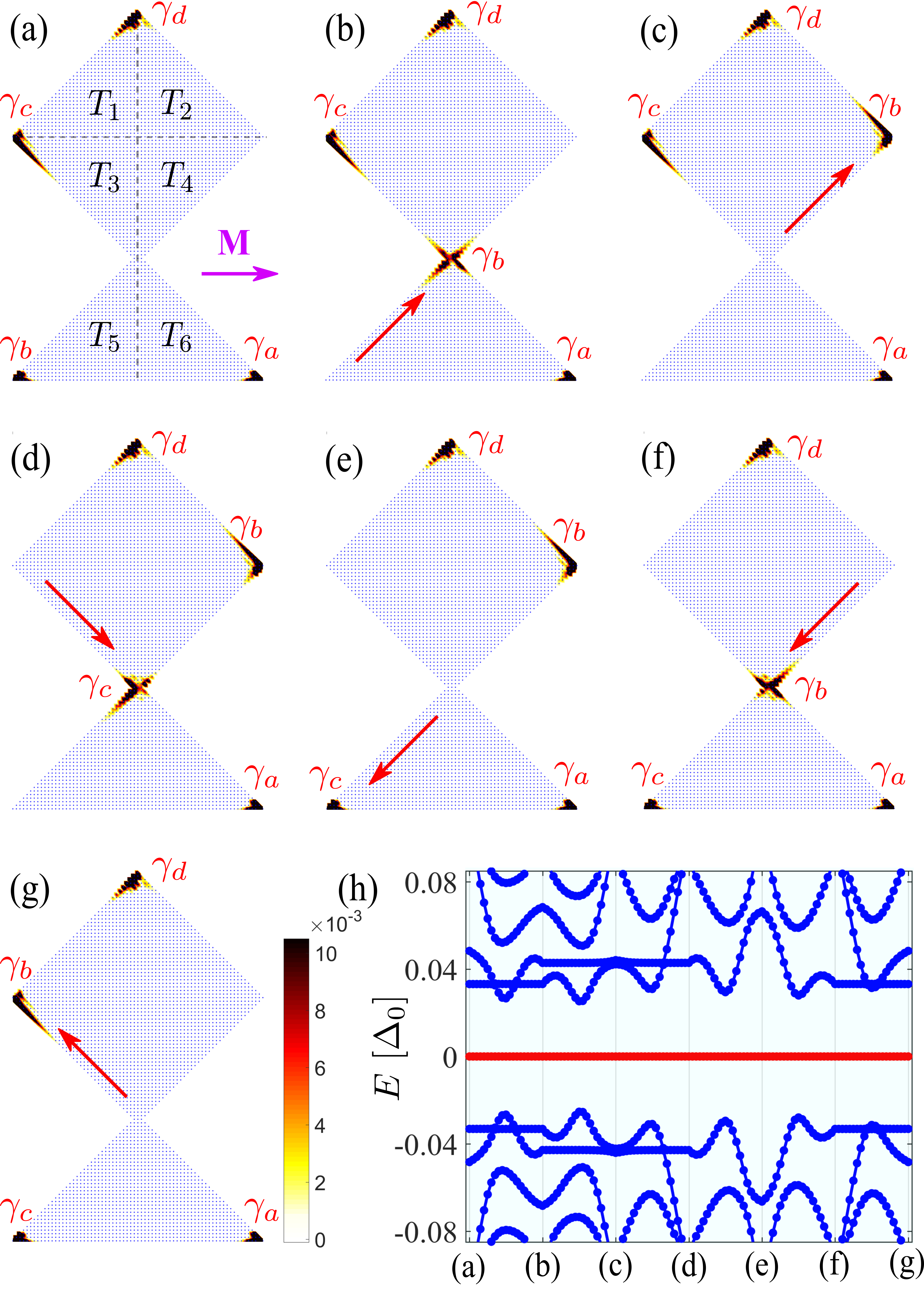}

\caption{Numerical simulation of the exchange of $\gamma_{b}$ and $\gamma_{c}$.
Seven subsequent snapshots show the positions of MBSs at (a) $\bm{\mu}_{345}\equiv(\mu_{3},\mu_{4},\mu_{5})=(\mu_{\text{d}},\mu_{\text{d}},\mu_{\text{u}})$,
(b) $\bm{\mu}_{345}=(\mu_{\text{d}},\mu_{\text{d}},\mu_{\text{d}})$,
(c) $\bm{\mu}_{345}=(\mu_{\text{d}},\mu_{\text{u}},\mu_{\text{d}})$,
(d) $\bm{\mu}_{345}=(\mu_{\text{u}},\mu_{\text{u}},\mu_{\text{d}})$,
(e) $\bm{\mu}_{345}=(\mu_{\text{u}},\mu_{\text{u}},\mu_{\text{u}})$,
(f) $\bm{\mu}_{345}=(\mu_{\text{u}},\mu_{\text{d}},\mu_{\text{u}})$,
and (g) $\bm{\mu}_{345}=(\mu_{\text{d}},\mu_{\text{d}},\mu_{\text{u}})$.
(h) Energy spectrum of the system during the exchange process. $\mu_{1}=\mu_{6}=\mu_{\text{u}}$,
$\mu_{2}=\mu_{\text{d}}$ \textcolor{black}{and other parameters are
the same as Fig.\ 3 of the main text.}}

\label{fig:braiding-bc}
\end{figure}

\begin{figure}[th]
\includegraphics[width=1\columnwidth]{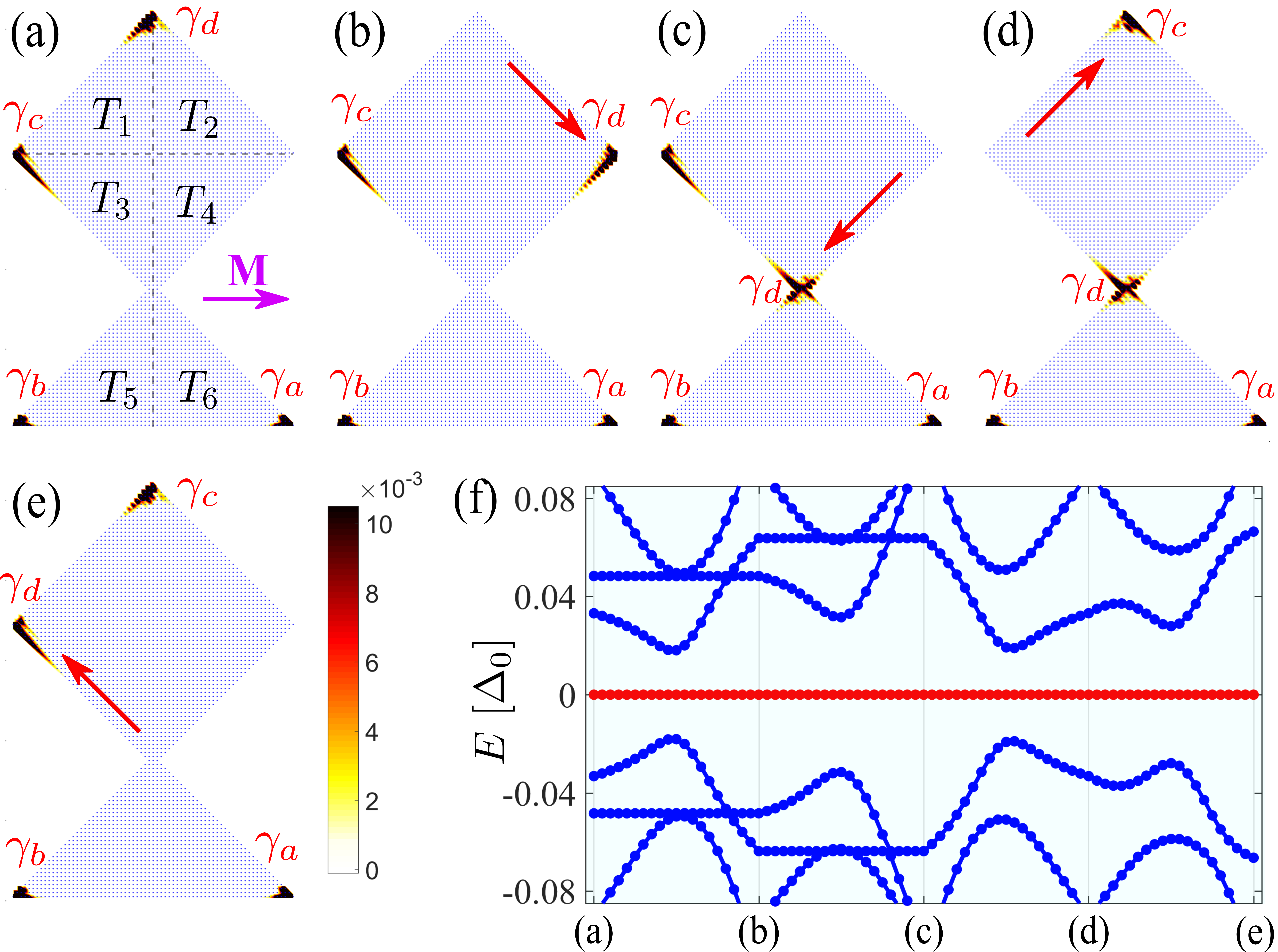}

\caption{Numerical simulation of the exchange of $\gamma_{c}$ and $\gamma_{d}$.
Five subsequent snapshots show the positions of MBSs at (a) $\bm{\mu}_{1234}\equiv(\mu_{1},\mu_{2},\mu_{3},\mu_{4})=(\mu_{\text{u}},\mu_{\text{d}},\mu_{\text{d}},\mu_{\text{d}})$,
(b) $\bm{\mu}_{1234}=(\mu_{\text{u}},\mu_{\text{u}},\mu_{\text{d}},\mu_{\text{d}})$,
(c) $\bm{\mu}_{1234}=(\mu_{\text{u}},\mu_{\text{u}},\mu_{\text{d}},\mu_{\text{u}})$,
(d) $\bm{\mu}_{1234}=(\mu_{\text{d}},\mu_{\text{u}},\mu_{\text{d}},\mu_{\text{u}})$,
and (e) $\bm{\mu}_{1234}=(\mu_{\text{d}},\mu_{\text{u}},\mu_{\text{u}},\mu_{\text{u}})$.
(f) Energy spectrum of the system during the process. $\mu_{5}=\mu_{6}=\mu_{\text{u}}$
\textcolor{black}{and other parameters are the same as Fig.\ 3 of
the main text.}}

\label{fig:braiding-cd}
\end{figure}

In Figs.\ \ref{fig:braiding-bc} and \ref{fig:braiding-cd}, we present
the numerical simulations of the exchanges of the MBS pairs, $\gamma_{b}\leftrightarrow\gamma_{c}$
and $\gamma_{c}\leftrightarrow\gamma_{d}$, respectively. These two
exchanges, together with that of $\gamma_{a}\leftrightarrow\gamma_{b}$
(which is presented in the main text), generate the whole braid group
of the four MBSs. For the exchange of $\gamma_{b}\leftrightarrow\gamma_{c}$,
we turn the chemical potentials in the following successions: (i)
$\mu_{5}=\mu_{\text{u}}\rightarrow\mu_{\text{d}}$; (ii) $\mu_{4}=\mu_{\text{d}}\rightarrow\mu_{\text{u}}$;
(iii) $\mu_{3}=\mu_{\text{d}}\rightarrow\mu_{\text{u}}$; (iv) $\mu_{5}=\mu_{\text{d}}\rightarrow\mu_{\text{u}}$;
(v) $\mu_{4}=\mu_{\text{u}}\rightarrow\mu_{\text{d}}$ ; and (vi)
$\mu_{3}=\mu_{\text{u}}\rightarrow\mu_{\text{d}}$. For the exchange
of $\gamma_{c}\leftrightarrow\gamma_{d}$, we tune (i) $\mu_{2}=\mu_{\text{d}}\rightarrow\mu_{\text{u}}$;
(ii) $\mu_{4}=\mu_{\text{d}}\rightarrow\mu_{\text{u}}$; (iii) $\mu_{1}=\mu_{\text{u}}\rightarrow\mu_{\text{d}}$
and (iv) $\mu_{3}=\mu_{\text{d}}\rightarrow\mu_{\text{u}}$. The corresponding
energy spectra are given in Figs.\ \ref{fig:braiding-bc}(h) and
\ref{fig:braiding-cd}(h). The MBSs (red flat bands) are protected
from excited modes (blue bands) by an excitation gap.

\section{Measurement of Majorana qubits via quantum dots\label{sec:Measurement-of-Majorana}}

\begin{figure}[bh]
\includegraphics[width=0.95\columnwidth]{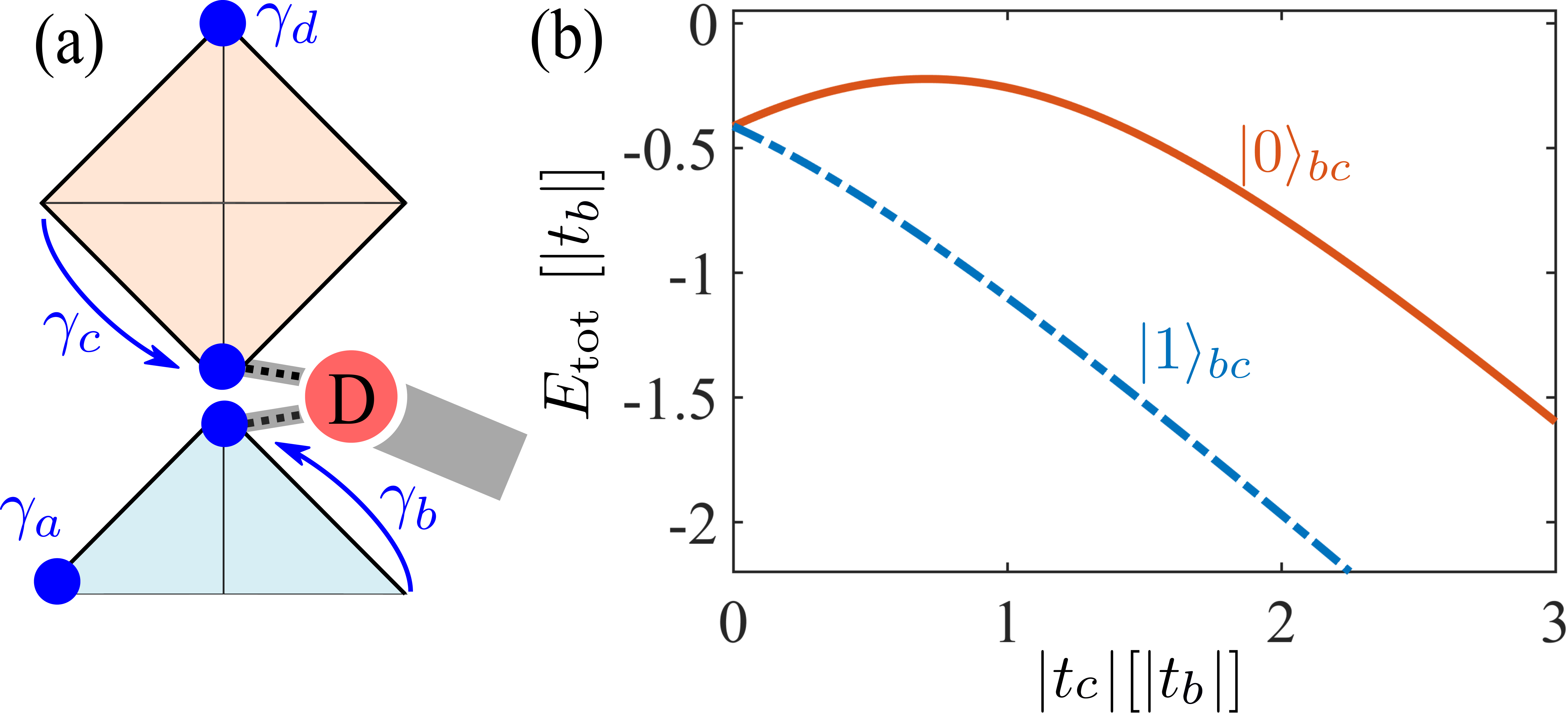}

\caption{(a) Schematics for probing the Majorana qubit via quantum dots. (b)
Perturbed ground-state energy of the quantum dot coupled to the MBSs
as a function of the coupling strength $|t_{c}|$. For concreteness,
we assume that the single energy level $\varepsilon$ of the dot is
empty. The solid and dashed curves correspond to the two qubit states
$P_{bc}|0\rangle_{bc}=|0\rangle_{bc}$ and $P_{bc}|1_{bc}\rangle=-|1_{bc}\rangle$.
$\varepsilon=3|t_{b}|$ and $\Phi_{bc}=\pi/2$ are used for the plotting.}

\label{fig:QDmeasure-braiding}
\end{figure}

In this section, we briefly discuss the qubit measurement by using
quantum dots. To do so, we turn off the connection at the junction
(e.g., by applying an external gate which generates a large barrier
potential) and move two measured MBSs, say again $\gamma_{b}$ and
$\gamma_{c}$, to the two disconnected vertices, respectively, as
sketched in Fig.\ \ref{fig:QDmeasure-braiding}(a). We assume a quantum
dot (in red) nearby with a single energy level $\varepsilon$ and
couple it elastically to $\gamma_{b}$ and $\gamma_{c}$ with coupling
amplitudes $t_{b}$ and $t_{c}$, respectively. In the Coulomb blockade
regime, the perturbed ground-state energy $E_{\text{tot}}$ of the
dot depends on the total fermion parity of the dot and two MBSs \citep{Flensberg11PRL},
i.e., $E_{\text{tot}}=\varepsilon/2-(\varepsilon^{2}/4+|t_{b}^{2}|+|t_{c}^{2}|+2\pm|t_{b}t_{c}|\sin\Phi_{bc}){}^{1/2}$,
where $\text{\ensuremath{\pm}}$ corresponds to the fermion parity
associated with the two MBSs $\gamma_{b}$ and $\gamma_{c}$, respectively,
and $\Phi_{bc}=2\text{Arg}(t_{b}/t_{c})$ depends on $\Phi_{0}$.
Suppose that the occupancy of the dot is known, then, the parity dependent
energy $E_{\text{tot}}$ {[}see Fig.\ \ref{fig:QDmeasure-braiding}(b){]}
could provide an alternative method deduce the qubit states.\\

\begin{figure}[tp]
\includegraphics[width=0.95\columnwidth]{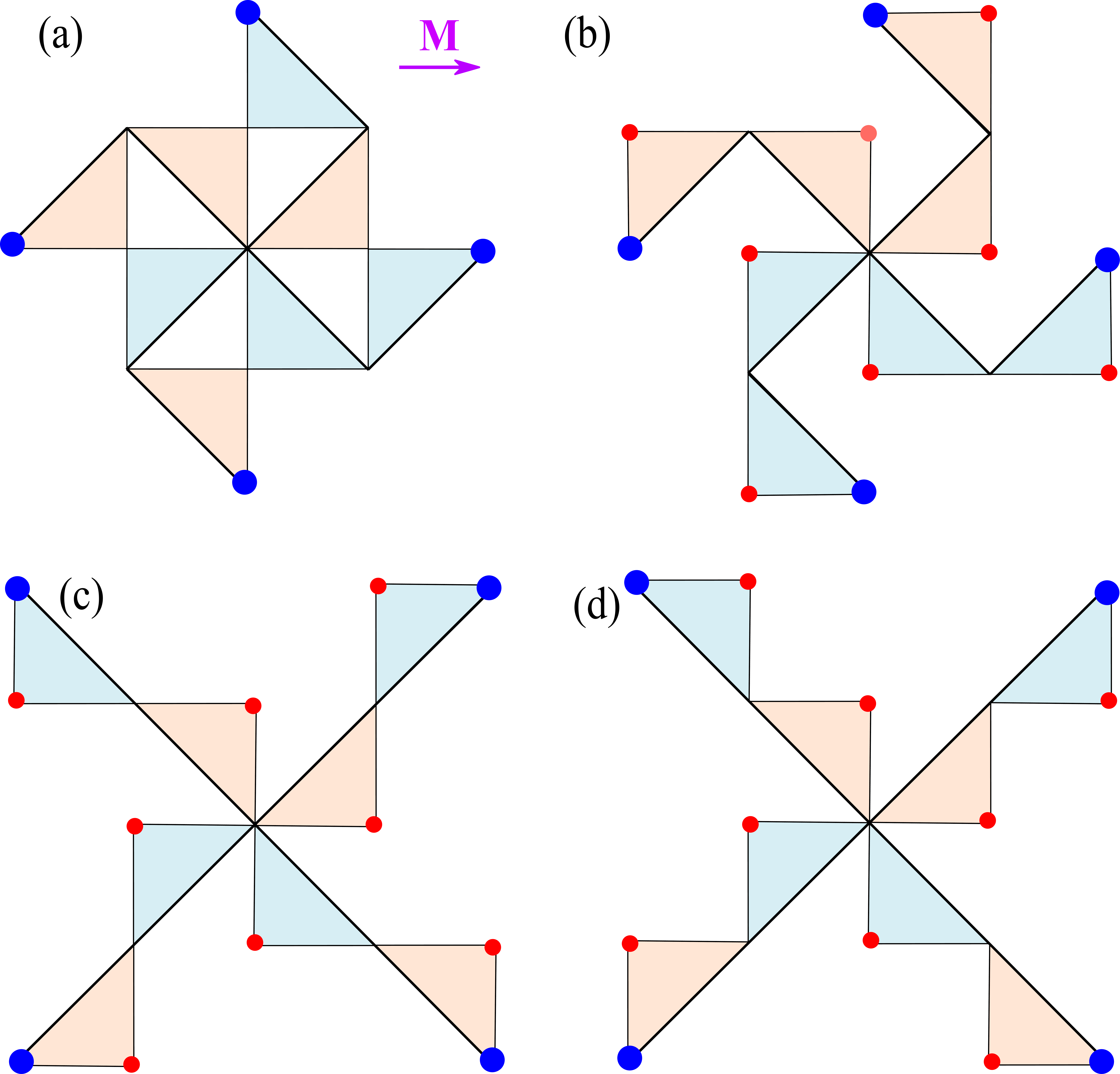}

\caption{Networks for MBS manipulation. The light blue and yellow color distinguish
between two different phases of pairing potentials in the islands.
The blue and red dots indicate the active and inactive MBSs, respectively.
These setups are also scalable by adding more IRTIs to the ``legs''.}

\label{fig:other-setups}
\end{figure}

\section{robustness with respect to chemical potential inhomogeneities within
individual IRTIs\label{sec:robustness-with-respect}}

\begin{figure*}[tp]
\includegraphics[width=0.95\textwidth]{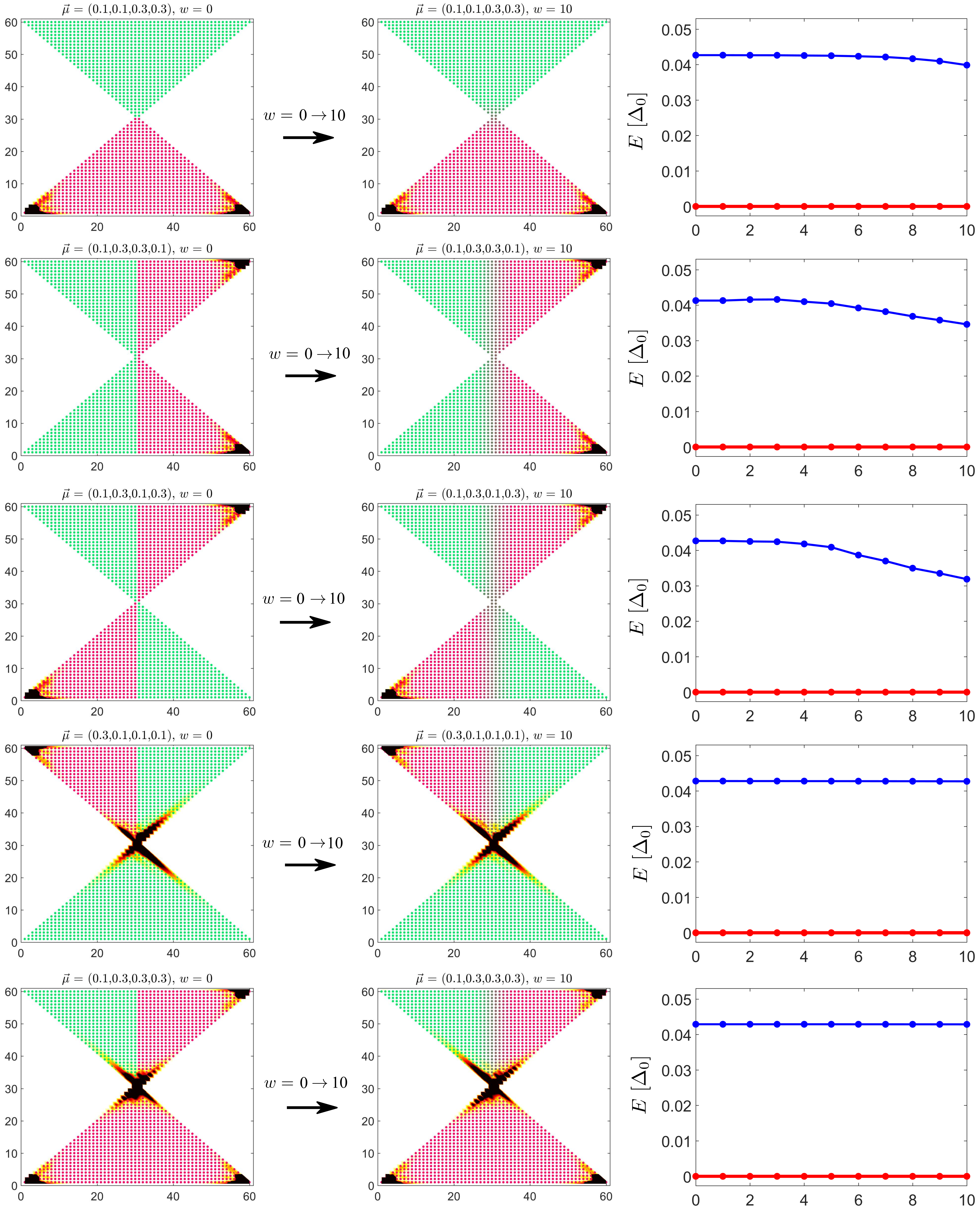}

\caption{Results in the presence of inhomogenous chemical potentials within
individual IRTIs. The chemical potentials in the four triangles $\protect\overrightarrow{\mu}\equiv(\mu_{1},\mu_{2},\mu_{3},\mu_{4})$
(in units of $0.5m_{0}$) and the parameter $w$ are given in the
titles of the panels in the first and second columns. The green-red
color indicates the chemical potential on each lattice site (with
the green limit for $\mu=\mu_{\text{d}}$ while the magenta limit
for $\mu=\mu_{\text{u}}$). The panels in the third rows correspond
to the energy spectra as functions of $w$. We can see clearly that
increasing $w$ barely change the results. The MBSs stay robustly
at the vertices and at zero energy. All the other parameters for all
panels are the same as those in Fig.\ \ref{fig:localization-length}.}

\label{fig:inhomogenous}
\end{figure*}

In all above simulations, we assume that the chemical potential is
homogeneous in each individual IRTI. In realistic situations, however,
variations of chemical potential are likely, especially, around the
boundary connecting to adjacent IRTIs. Therefore, in this section,
we study the effect of such chemical potential inhomogeneity within
each triangle. We focus on the junction between four triangles, considering
all possible configurations of the chemical potentials. For simplicity,
we use the parameter $w$ to model the width over which the chemical
potential interpolates linearly between the values of adjacent triangles.
Typical results are displayed in Fig.\ \ref{fig:inhomogenous}. In
the left and center plots, the color indicates the chemical potential
on each lattice site. The wavefunctions of MBSs are plotted on top.
In the right plots, we show the low-energy part of the spectra as
a function of $w$. As we can see, no important changes appear up
to $w=10$ (the size of the triangles is 30 sites on the short sides).
The wavefunction of MBSs seems to be barely affected, as well as the
energy spectrum. Note that deviations from perfect degeneracy at zero
energy can be exponentially suppressed by increasing the system size.

\section{Other alternative setups for braiding Majorana bound states\label{sec:Other-setups}}

In this section, we present some other simple network examples for
Majorana manipulation, as shown in Fig.\ \ref{fig:other-setups}.
Different from the one shown in the main text, we construct these
networks by connecting the IRTIs only at the vertices. But similarly,
two different phases of pairing potential (indicated by the cyan and
yellow colors) are applied in a strip form. The ferromagnetic of antiferromagnetic
order (${\bf M})$ in the bulk is fixed and uniform. By locally controlling
the chemical potentials in the IRTIs, we are not only able to increase
or reduce the number of MBS pairs but also to control their positions.
The Majorana qubits are also measurable via the Josephson effect or
adjacent quantum dots, in a similar way we discussed before. Note
that because the MBSs at the right-angle vertices of IRTIs are immobile,
we can design our setups without connecting the right-angle vertices
of IRTIs. In this case, extra MBSs appear at the unconnected right-angle
vertices. However, they are ``inactive'' when performing braiding
or measuring operations.


\begin{thebibliography}{72}
\bibitem [{Not()}]{NoteM}%
  \BibitemOpen
  \href@noop {} {\bibinfo  {journal} {We note, however, that our main results discussed below also apply to other cases, for instance, AFM with $H_M=M_0\tau_zs_x\sigma_x$}\ }\BibitemShut {NoStop}%
\bibitem [{Not()}]{NoteN}%
  \BibitemOpen
  \href@noop {} {\bibinfo  {journal} {The pairing potential in
  FeTe$_{1-x}$Se$_x$ may be $s_\pm$-wave. In this case, we should replace
  $\Delta_0\rightarrow\Delta_0+ \Delta_2(\cos k_x+\cos k_y)$ in the bulk
  Hamiltonian. The momentum correction in the pairing potential modifies the
  effective pairing potential on the boundary to
  $\Delta_0\rightarrow\Delta_0+\Delta_2 m_0/m$ and thus does not change the
  results qualitatively}\ }\BibitemShut {NoStop}%
\end{thebibliography}
\end{document}